\documentclass[letterpaper,twocolumn,10pt]{article}
\usepackage{arXiv}
\usepackage{mathptmx}
\usepackage{lipsum}
\usepackage{xcolor}

\usepackage{booktabs}
\usepackage[labelfont={small,bf},textfont={small,bf}]{caption}
\usepackage{subcaption}
\usepackage{cite}

\usepackage{multirow,colortbl}
\usepackage[noend, ruled]{algorithm2e}
\usepackage{mathtools,amsmath}

\usepackage{amssymb} %
\usepackage{amsfonts}
\usepackage{dsfont}
\usepackage{float}
\usepackage{enumitem}
\setlist[itemize]{leftmargin=*}

\usepackage{makecell}
\usepackage{tcolorbox}
\usepackage[absolute]{textpos}
\usepackage{tcolorbox}

\newcommand{\mypara}[1]{\smallskip\noindent\textbf{#1.}\xspace}

\definecolor{mygray}{gray}{.9}

\newcommand{\sysname}{\textsc{Face-Auditor}\xspace}
\newcommand{\siamese}{SiameseNet\xspace}
\newcommand{\proto}{ProtoNet\xspace}
\newcommand{\relation}{RelationNet\xspace}

\newcommand{\model}{\ensuremath{\mathcal{M}}\xspace}
\newcommand{\dset}{\ensuremath{\mathcal{D}}\xspace}

\newcommand{\feat}{\ensuremath{\mathcal{E}}\xspace}
\newcommand{\auditfeat}{\ensuremath{\chi}\xspace}

\newcommand{\support}{\ensuremath{\mathbb{S}}\xspace}
\newcommand{\query}{\ensuremath{\mathbb{Q}}\xspace}
\newcommand{\probe}{\ensuremath{\mathbb{P}}\xspace}

\usepackage{etoolbox}
\makeatletter
\patchcmd{\hyper@makecurrent}{%
    \ifx\Hy@param\Hy@chapterstring
        \let\Hy@param\Hy@chapapp
    \fi
}{%
    \iftoggle{inappendix}{%
        \@checkappendixparam{chapter}%
        \@checkappendixparam{section}%
        \@checkappendixparam{subsection}%
        \@checkappendixparam{subsubsection}%
        \@checkappendixparam{paragraph}%
        \@checkappendixparam{subparagraph}%
    }{}%
}{}{\errmessage{failed to patch}}

\newcommand*{\@checkappendixparam}[1]{%
    \def\@checkappendixparamtmp{#1}%
    \ifx\Hy@param\@checkappendixparamtmp
        \let\Hy@param\Hy@appendixstring
    \fi
}
\makeatletter

\newtoggle{inappendix}
\togglefalse{inappendix}

\apptocmd{\appendix}{\toggletrue{inappendix}}{}{\errmessage{failed to patch}}

\begin{document}

\begin{textblock}{16}(3,1.5)
To appear in the 32nd USENIX Security Symposium, August 2023, Anaheim, CA, USA
\end{textblock}

\date{}

\title{\Large \bf \sysname: Data Auditing in Facial Recognition Systems}
\author{
Min Chen\textsuperscript{1}\ \ \
Zhikun Zhang\textsuperscript{1,2}\thanks{Corresponding authors.}\ \ \
Tianhao Wang\textsuperscript{3}\ \ \
Michael Backes\textsuperscript{1}\ \ \ 
Yang Zhang\textsuperscript{1}\textsuperscript{{$\ast$}}
\\
\\
\textsuperscript{1}\textit{CISPA Helmholtz Center for Information Security} \ \ \ 
\\
\textsuperscript{2}\textit{Stanford University} \ \ \

\textsuperscript{3}\textit{University of Virginia} \ \ \
}
\maketitle
\pagestyle{plain}

\begin{abstract}
Few-shot-based facial recognition systems have gained increasing attention due to their scalability and ability to work with a few face images during the model deployment phase.
However, the power of facial recognition systems enables entities with moderate resources to canvas the Internet and build well-performed facial recognition models without people's awareness and consent.
To prevent the face images from being misused, one straightforward approach is to modify the raw face images before sharing them, which inevitably destroys the semantic information, increases the difficulty of retroactivity, and is still prone to adaptive attacks.
Therefore, an auditing method that does not interfere with the facial recognition model's utility and cannot be quickly bypassed is urgently needed.

In this paper, we formulate the auditing process as a user-level membership inference problem and propose a complete toolkit \sysname that can carefully choose the probing set to query the few-shot-based facial recognition model and determine whether any of a user's face images is used in training the model.
We further propose to use the similarity scores between the original face images as reference information to improve the auditing performance.
Extensive experiments on multiple real-world face image datasets show that \sysname can achieve auditing accuracy of up to $99\%$.
Finally, we show that \sysname is robust in the presence of several perturbation mechanisms to the training images or the target models.\footnote{The source code of our experiments can be found at \url{https://github.com/MinChen00/Face-Auditor}.}
\end{abstract}

\section{Introduction}
\label{section:introduction}
Facial recognition is widely used to perform identification~\cite{WD21,TYRW14,SKP15,GHBNFGBKKZBG19}.
Modern facial recognition system utilizes machine learning models to determine whether a face image being verified belongs to the authorized users (the complete system also includes other components like face detection~\cite{YLLT16}, liveness detection~\cite{NLM16}, etc).
In the training phase, a facial recognition model takes in multiple images for each user (e.g., from different angles) in advance. 
In the identification phase, the facial recognition model compares the image being examined with the pre-existing pictures to determine whether it belongs to the authorized users and (if yes) to which authorized user this image belongs.
More recently, few-shot learning~\cite{KZS15,SSZ17,SYZXTH18,GZZCLL20,FYTSC21} dominate the traditional learning in facial recognition systems because it requires only \textit{a few} ``anchor'' face images from the authorized users.

With the power of facial recognition systems, entities with moderate resources can canvas the Internet for face images and build well-performed facial recognition models without people's awareness and consent.
For example, \url{clearview.ai} reveals that a private company has collected 3 billion online face images and trained a powerful model capable of recognizing millions of citizens.
Such kinds of misuse of facial recognition systems are potentially disastrous~\cite{MAP21} and infringe the privacy laws such as European Union's General Data Protection Regulation (GDPR).
GDPR states that \textit{the personal data must only be processed if the individual has given explicit consent} (Article 6(1)(a)), and \textit{the processing of personal data must be lawful, fair, and transparent} (Article 5(1)(a))~\cite{GDPR}.
This means that if the third parties want to use the data owner's face images, they need to obtain consent from the data owner and inform the data owner how their face images are processed.
Sharing personal data online typically implies that the data owners are willing to share their data with the public for social or promotion purposes.
However, this does not grant others the right to misuse the data for unconsent purposes, particularly in commercial activities.

To prevent face images from being misused, one straightforward method is to modify the raw face images before uploading them to the Internet, such as distorting the face images~\cite{LL19}, producing adversarial patches~\cite{TRG19}, or adding imperceptible pixel-level cloaks~\cite{SWZLZZ20}.
However, these approaches inevitably destroy the semantic information of the face images and also increase the difficulty of retroactivity.
Also, researchers have argued that such defenses can be bypassed by newer technologies~\cite{RHCT22}, which leads to an endless arms race between the attacker and defender.

\subsection{Our Contributions}
In this paper, we take a different angle by advocating a responsible \textit{auditing} approach that enables normal users to detect whether their private face images are being used to train a facial recognition system.
This approach provides users with evidence in claiming proprietary of their face images.
Furthermore, it complies with data privacy protection regulations such as GDPR, which gives users the right to know how their data is processed.
If data owners do not want any entity to use their face images to train the facial recognition system, they can use \sysname to audit if their face images are being used. 
If they find their face images were used without their consent, the data owners can take legal action against the model developer in accordance with GDPR regulations.

Concretely, we propose \sysname to determine whether a target user's face images were used to train a facial recognition model.
The underlying problem can be formalized as \emph{user-level} membership inference.
Different from classic \textit{sample-level} membership inference, which detects whether a specific sample was used for training the target model, user-level membership inference aims to determine whether any of a target user's data was used to train the model.
Here, the auditor has a set of samples (images) of a target user, and these samples are not necessarily used to train the target facial recognition model to claim/predict membership.

\mypara{Methodology}
We discuss the details of the technical challenges and provide a systematic analysis of how we address each of them in \autoref{subsection:technique_novelty}. 
Briefly, to obtain an auditing model that works broadly, we adopt the well-established yet comprehensive {\it shadow model paradigm} that aims to mimic the behavior of the target facial recognition model:
We assume the auditor can access an auxiliary dataset to train the shadow model.
In a more restricted and practical scenario, the auxiliary data does not need to share the same distribution as the target model's training dataset.
To achieve the goal of user-level auditing, \sysname accepts a set of target face images as input and outputs a binary indicator of a member user or non-member user.
To cope with the few-shot learning paradigm, given the target face images, we supplement a set of anchor face images to form a {\it probing set}. 
We then use the probing set to query the target facial recognition system and generate ``posteriors'' (a sequence of similarity scores from the target face to the anchor faces) as the features for \sysname.
Here, the target face image is not necessarily used to train the target model.
To further improve the auditing performance, we propose using reference information to strengthen the auditing feature, which can be calculated by comparing the original target face and anchor images.

\mypara{Evaluation}
We conduct experiments on three representative few-shot learning algorithms and four human face datasets to illustrate the effectiveness of \sysname.
The experimental results show that \sysname can achieve up to $99\%$ auditing accuracy on the \siamese model. 
We observe that when the target model has high representation capability, it is more difficult to audit.
Furthermore, we conduct experiments to validate that adding the reference information of the original image can effectively improve the auditing performance.
For instance, after introducing reference information to audit the \relation model, we achieve $72\%$ accuracy improvement.

\mypara{Robustness}
In practice, the target model might be equipped with different obfuscation techniques to preserve the privacy of the training data~\cite{ACGMMTZ16,LL19,TRG19,SWZLZZ20,JSBZG19}.
Therefore, we conduct experiments to validate the robustness of \sysname when the training images or the target models are protected.
Concretely, we consider three representative privacy-preserving mechanisms in a general ML model pipeline: Input perturbation (perturb the training images), training perturbation (perturb the training gradient by enforcing differential privacy), and output perturbation (perturb the output similarity scores).
We also show the robustness of \sysname under an adaptive attack setting, where the target model's output is perturbed specifically to evade the auditing from \sysname.
We observe that the performance of \sysname only slightly drops, which indicates the robustness of \sysname.

In summary, our contributions are four-fold:
\begin{itemize}
    \item We take the first step to investigating the auditing approach that enables ordinary users to detect whether their private face images are being used to train a facial recognition system when only similarity metric information is accessible.
    \item We carefully design the probing set for querying the target facial recognition model and propose using multiple metrics to construct the reference feature to enhance the auditing performance.
    \item We systematically evaluate the factors that affect the auditing performance and highlight some design oracles for an effective auditor.
    \item In practice, an advanced adversary might be aware of the existence of the auditor and try to evade the detection of their misuse. 
    Therefore, we investigate the robustness of \sysname when the training images or the target models are protected by different defense mechanisms.
\end{itemize}

\subsection{Organization}
We discuss the related work in \autoref{section:related_work}.
In \autoref{section:preliminary}, we introduce the facial recognition system and the application of few-shot learning.
In \autoref{section:audit_method}, we formulate the auditing process as user-level membership inference and depict the design details of \sysname.
We perform experiments in \autoref{section:evaluation} to illustrate the effectiveness of \sysname. 
We discuss the ethical consideration and practical impacts of \sysname in \autoref{section:discussion} and conclude the paper in \autoref{section:conclusion}.

\section{Related Work}
\label{section:related_work}

\mypara{Privacy of Facial Recognition Systems}
With the proliferation of facial recognition systems, their privacy issues have attracted increasing attention~\cite{TZL21,CGTFJB21,WSZZ22}.
To protect users' privacy, one strategy is to make the face images difficult for a facial recognition system to recognize by relying on adversarial examples~\cite{ESK21,CGTFJB21}.
Sharif et al.~\cite{SBBR16} show that adding specially printed glasses can cause the wearer to be misidentified.
Komkov et al.~\cite{KP20} propose to add carefully computed adversarial stickers on a hat to reduce its wearer's likelihood of being recognized.
Others propose to add adversarial patches to make it difficult for facial recognition systems to recognize the user as a person in an image~\cite{TRG19,WLDG20}.
An alternative is to evade the facial recognition models by poisoning their training samples.
One representative method is Fawkes~\cite{SWZLZZ20}.
However, these approaches can inevitably destroy the semantic information of the face images and are still vulnerable to advanced adversaries~\cite{RHCT22}.

\mypara{Sample-level Membership Inference}
Previous studies on membership inference attacks mainly focus on sample-level membership inference~\cite{SSSS17,SZHBFB19,MSCS19,SS19,CZWBHZ21}.
The first membership inference attack was proposed by Shokri et al.~\cite{SSSS17}, which uses shadow models to mimic the target model’s behavior and generate training data for the attack model.
Salem et al.~\cite{SZHBFB19} gradually removed the assumptions of~\cite{SSSS17} by proposing three different attack methods.
More recently, membership inference has been extensively investigated in various ML models and tasks, such as federated learning~\cite{MSCS19},
natural language processing~\cite{SS19}, 
and neural architecture search~\cite{HZSBLZ22}. 

To mitigate the threat of membership inference, a plethora of defense mechanisms have been proposed.
These defenses can be classified into Reducing overfitting, perturbing posteriors, and adversarial training.
There are several ways to reduce overfitting in the machine learning field, such as $\ell_2$-regularization~\cite{SSSS17}, dropout~\cite{SZHBFB19}, and model stacking~\cite{SZHBFB19}.
In~\cite{LLR21}, the authors proposed to explicitly reduce the overfitting by adding to the training loss function a regularization term, which is defined as the difference between the output distributions of the training set and the validation set.
Jia et al.~\cite{JSBZG19} proposed a posterior perturbation method inspired by adversarial examples.
Nasr et al.~\cite{NSH18} proposed an adversarial training defense to train a secure target classifier.
During the training of the target model, a defender’s attack model is trained simultaneously to launch the membership inference attack. 
The optimization objective of the target model is to reduce the prediction loss while minimizing the membership inference attack accuracy. 

\mypara{User-level Membership Inference}
Compared to the sample-level membership inference, the user-level inference is less investigated.
The first user-level membership inference was proposed by Song et al.~\cite{SS19} for the natural language models, including next-word prediction, neural machine translation, and dialog generation.
They design and evaluate a black-box auditing method that can detect, with very few queries to a model, if a particular user's texts were used to train it.
Miao et al.~\cite{MXCPZZKX21} then investigate the user-level membership inference against the automatic speech recognition model. 
Note that Audio-Auditor feeds multiple audios of the target user to the target model independently and obtains multiple transcriptions.
The auditing features are constructed by combining the input audio, input transcriptions, output transcriptions, and their statistical information. 
\sysname needs to carefully design the probing set and combine the similarity scores as the auditing feature. 
We also introduce image-level similarity as reference information to improve auditing performance. 
Furthermore, \sysname does not need access to the exact face images used to train the target model; instead, it only needs to take a few face images of the target user. While Audio-Auditor does not have this property (at least in their experiments). 

\begin{figure*}[!tpb]
    \centering
    \begin{subfigure}{2\columnwidth}
    \includegraphics[width=\textwidth]{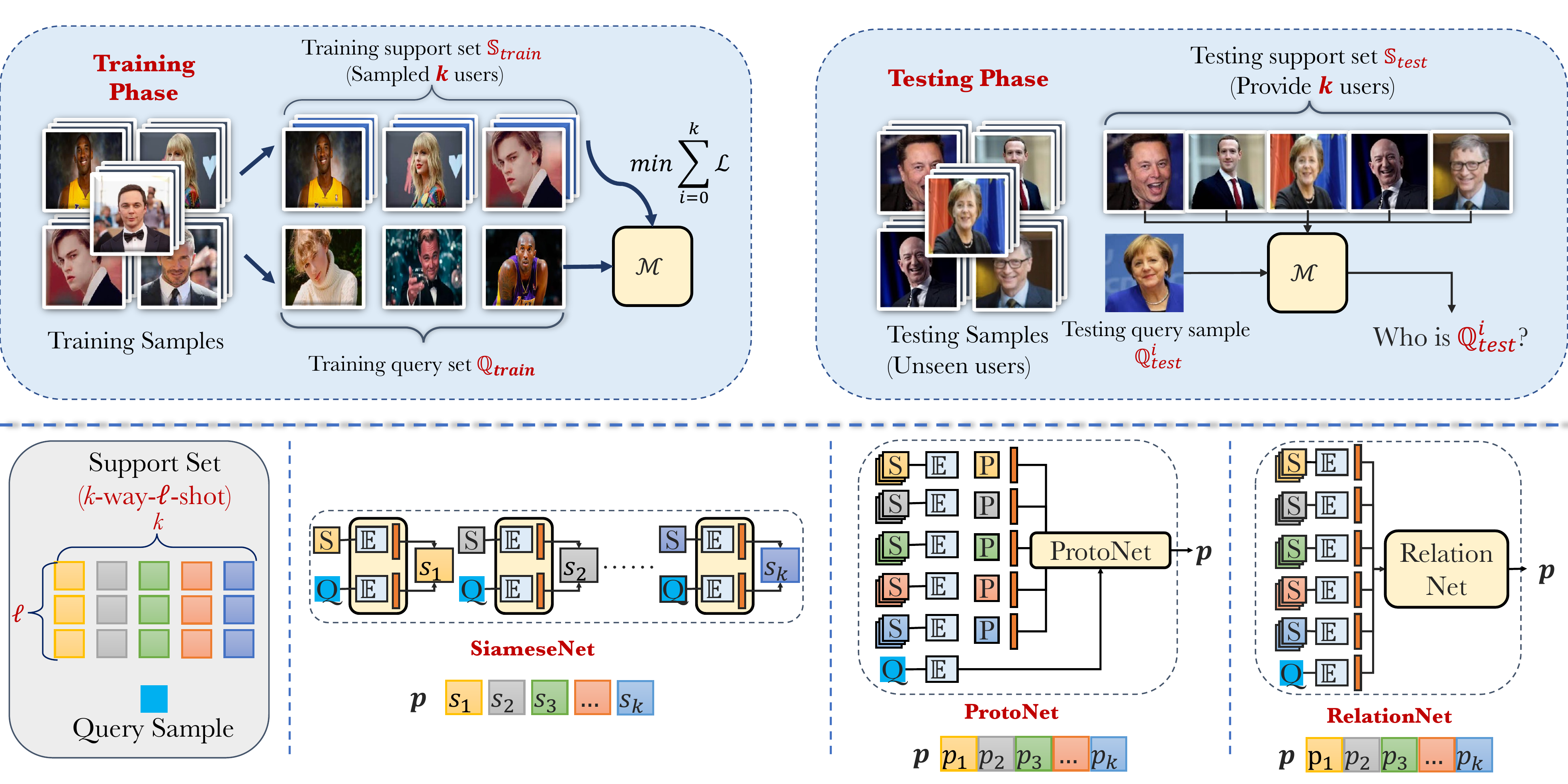}
    \label{subfigure:model_structure}
    \end{subfigure}
    \vspace{-0.6cm}
    \caption{
    Illustration of the metric-based few-shot facial recognition models.
    The algorithm consists of both the training phase and the testing phase.
    The objective of the training phase is to train a feature extractor \feat to make the images in the query set \query close (in terms of the embeddings) to the images in the support set \support if they come from the same user, and make them far away when they are from different users.
    In the testing phase, given a query image, we predict its identity label as the closest user in the support set. 
    A \siamese takes image pairs as input and outputs a similarity score. 
    Both \proto and \relation takes as input the support set simultaneously and outputs a posteriors vector. 
    \proto measures the similarity by Euclidean distance while \relation explicitly learns a trainable relation module.}
    
    \label{figure:model_structure}
\end{figure*}

The most relevant study with \sysname is Li et al.~\cite{LRL22}.
While we both focus on user-level membership inference against metric learning models and share the same intuition that the images from a member user tend to be closer to each other in the latent space.
However, our work differs from Li et al.~\cite{LRL22} in multiple aspects.
First, the \textbf{threat model is different}. 
Li et al. assume the adversary can access the embeddings of the target model, while \sysname can only obtain the similarity scores, which is more practical in real-world facial recognition systems and more challenging to design the inference/auditing model.
Second, the \textbf{feature design is different}.
Li et al. feed all the face images of the target user to the embedding extractor and use the embedding distances of the input images as the attack feature.
On the other hand, \sysname carefully designs the probing set to query the target model and uses the similarity score as the basic auditing feature.
We further discover that the raw image similarity can serve as a decisive reference information that significantly increases the auditing performance.
Finally, the \textbf{application range is different}.
\sysname achieves good auditing performance for both simple models such as \siamese and complex models such as \proto and \relation, while Li et al. only achieve acceptable performance for \siamese. 
We refer the readers to see the detailed experiments in \autoref{app:embedding_results}. 

\mypara{Attacks for AI System Auditing}
Using attacks against machine learning as an auditing tool has been a growing trend in trustworthy AI~\cite{SS19,LWHSZBCFZ22}. 
``Desirable attacks''~\cite{APSK20} against ML, as an example, can be used for legitimate concerns like human rights and civil liberties. 
Determining whether a given image is present in a facial recognition database~\cite{FJR15} can help individuals determine whether they can bring a court case against the service provider. 
Model inversion ~\cite{FJR15} can detect potential bias decision-making in credit risk evaluation systems. 
Adversarial examples can be used as an obfuscation tool to make users less likely to be tracked~\cite{equalais} or re-identified~\cite{SWZLZZ20}. 
A similar notion of "subversive AI" adopts human-centered enhanced adversarial machine learning to evade algorithmic surveillance before publishing content online. 
Protective Optimization Technologies (POTs)~\cite{KOTG20} offer a more general terminology for repurposing the original system to enhance privacy, evade discrimination, or avoid surveillance from facial recognition systems.

\section{Preliminaries}
\label{section:preliminary}

\subsection{Facial Recognition System}
\label{subsection:facial_recognition_system}
The objective of the facial recognition system is to identify face images.
Formally, there is a pre-defined set of persons, which we call \textit{authorized users}.  
They each contribute multiple face images (which we call \textit{anchor images}) for the system to ``memorize'' them so that when a new face image comes, the system knows which (if any) authorized user this image corresponds to. 
A straightforward approach is to train a classification model with the anchor face images.
However, the classification model oftentimes requires a large amount of training data, while it is difficult to collect many face images from each authorized user in practice.
Furthermore, the set of authorized users often changes over time, for example, when new colleagues join or leave a company.
The classification model needs to be retrained when the set of authorized users changes. 
To address these challenges and improve the scalability of facial recognition systems, companies have turned to using few-shot learning techniques~\cite{WYKN20}.

\subsection{Few-shot Learning for Facial Recognition}
\label{subsection:fsl_models}
Few-shot learning is a machine learning paradigm that aims to obtain good learning performance given limited supervised information in the training set~\cite{CLKWH19}.
The high-level idea of few-shot learning is to exploit prior knowledge to help train, thus reducing the size of the actual training set.
An important branch of commonly used few-shot learning algorithms is based on \textit{metric learning}, which learns the similarity/relation (measured by some metric) among the images instead of (in the traditional classification problem) learning the mapping from an image to a specific label. 

\autoref{figure:model_structure} illustrates the general pipeline of metric-based few-shot learning algorithms, which consists of training and testing phases (also called the deployment phase in facial recognition systems).
The training phase takes a large, publicly available training dataset $\dset_{train}$ (which consists of samples for many classes) and runs in multiple iterations.
In each iteration, we construct a {\it support set} $\support_{train}$, which consists of randomly selected $k$ classes, each with $\ell$ samples, from $\dset_{train}$ (this is referred to as $k$-way-$\ell$-shot few-shot learning).
We also construct a \textit{query set} $\query_{train}$ similar to $\support_{train}$ by sampling from the same classes (note that \textit{query set} might be a bit confusing when used in training, but this is the standard terminology in few-shot learning).
Our goal is to train a feature extractor \feat so that the features (embeddings) of the images from $\support_{train}$ and $\query_{train}$ are optimized to be similar/close (in terms of some metric) if they belong to the same class, and dissimilar if they are from different classes. 
In the testing/deployment phase, we have a new support set $\support_{test}$ (anchor images from authorized users).
Since \feat has already been trained to perform well in distinguishing samples from different classes; given one query image, we predict it as the closest class in $\support_{test}$. 
As $\support_{test}$ is taken as input in the testing case, and we only care about similarities, it is easy to add/remove authorized users in few-shot learning.

Different metric-based few-shot algorithms vary in their strategies for making predictions conditioned on the support set.
In the following, we introduce several representative metric-based few-shot algorithms~\cite{WD21,TYRW14,KZS15}:

\mypara{Siamese Network (\siamese)~\cite{LWYLRS17}}
The most simple yet commonly used few-shot learning algorithm relies on the Siamese network, which inputs a pair of images and outputs their similarity score. 
It consists of a feature extractor \feat that learns the embedding of each image and a similarity metric (e.g., cosine similarity) that compares any two embeddings.
The objective is to train \feat so that the image pairs with the same label (positive pairs) have high similarity scores (in the embedding space), and image pairs with different labels (negative pairs) have low similarity scores.

As the \siamese is designed to learn the similarity between two images, it can be easily adapted to deal with the few-shot learning tasks.
Concretely, in the training phase, we pair the images from the support set $\support_{train}$ and the query set $\query_{train}$ one by one.
If the query image and the support image come from the same user, they form a positive pair; otherwise, they form a negative pair.
In the testing phase, given a query image $\query_{test}^{i}$, we compare it with all the images in the testing support set $\support_{test}$.
If the largest similarity score exceeds a predefined threshold, the query image belongs to the corresponding user; otherwise, the target image does not belong to any user.

\mypara{Prototypical Network (\proto)~\cite{SSZ17}}
\proto is specially designed for few-shot learning tasks.
It also contains a feature extractor \feat that transforms the images into embeddings.
Different from \siamese which takes pairs, \proto takes all the samples from the support set $\support_{train}$ simultaneously and compares the similarity between the query images in $\query_{train}$ and the support images in $\support_{train}$. 
For each class in the support set $\support_{train}$, we calculate the mean of the embeddings and generate a ``prototype''.
In total, we have $k$ prototypes in a $k$-way-$\ell$-shot few-shot learning task.
The objective is to train \feat to make the images in the query set $\query_{train}$ close to the prototype with the same user and far from the prototypes with different users.
The distance between the query embedding and the prototype is measured by \textit{Euclidean distance}. 

\mypara{Relation Network (\relation)~\cite{SYZXTH18}}
\relation shares a similar paradigm with \proto, which consists of a feature extractor \feat to transform the support set $\support$ into prototypes and the query set $\query$ into image embeddings.
It also aims to make the images in the query set $\query_{train}$ close to the prototype from the same user, and far from the prototypes from different users.
The main difference from \proto is that, instead of using Euclidean distance to measure the distance between query embedding and the prototype, it explicitly learns a trainable relation module, which typically consists of multiple stacked fully connected layers.

\section{Auditing Methodology}
\label{section:audit_method}

\subsection{Problem Statement}
\label{subsection:problem_statement}

\mypara{Auditing Goal}
We aim to determine whether any of the \textit{target user} $u$'s face images were used to train a target facial recognition model $\model_T$ (\textit{target model} for short).
We formulate this auditing process as a \emph{user-level} membership inference problem.
Formally, assume the target user $u$ has a set of face images $\mathcal{U}=\{u_1, u_2, \cdots, u_n\}$, the user-level membership inference aims to distinguish between $\mathcal{U} \cap \dset_{train}^T \neq \varnothing$ (member user) and $\mathcal{U} \cap \dset_{train}^T = \varnothing$ (non-member user), where $\dset_{train}^T$ is the training dataset of $\model_T$.
This is different from the classic \textit{sample-level} membership inference that aims to determine whether a specific face image was used to train the target model, i.e., $u_i \in \dset_{train}^T$ (member sample) or $u_i \not\in \dset_{train}^T$ (non-member sample). 

\mypara{Auditing Scenario}
Facial recognition systems are often trained by computer vision companies and sold to individual users or other companies for deployment. 
The model developer of the facial recognition system might collect face images from the Internet and misuse these face images without the data owners’ consent. 
Users who want to audit potential misuse of their face images could use \sysname as a privacy-auditing tool.

Note that \sysname is unnecessary to be trained by individuals. 
Alternatively, a third party with legal access to facial images (such as a qualified Auditing-ML-as-a-Service company, law enforcement, or government agency) can purchase the well-known facial recognition systems in the market and provides (free or charged) auditing services to individuals. 
By doing so, the third-party entity can ensure auditing accuracy and efficiency, making it more convenient for users who want to audit their face images. 
Individuals can quickly check if their face images are being used without their consent and take appropriate actions, such as reporting to the authorities or suing the model developer per the protection of privacy regulations~\cite{GDPR,CCPA,UKDPA}. 

\begin{figure*}[!t]
    \centering
    \begin{subfigure}{2\columnwidth}
    \includegraphics[width=\textwidth]{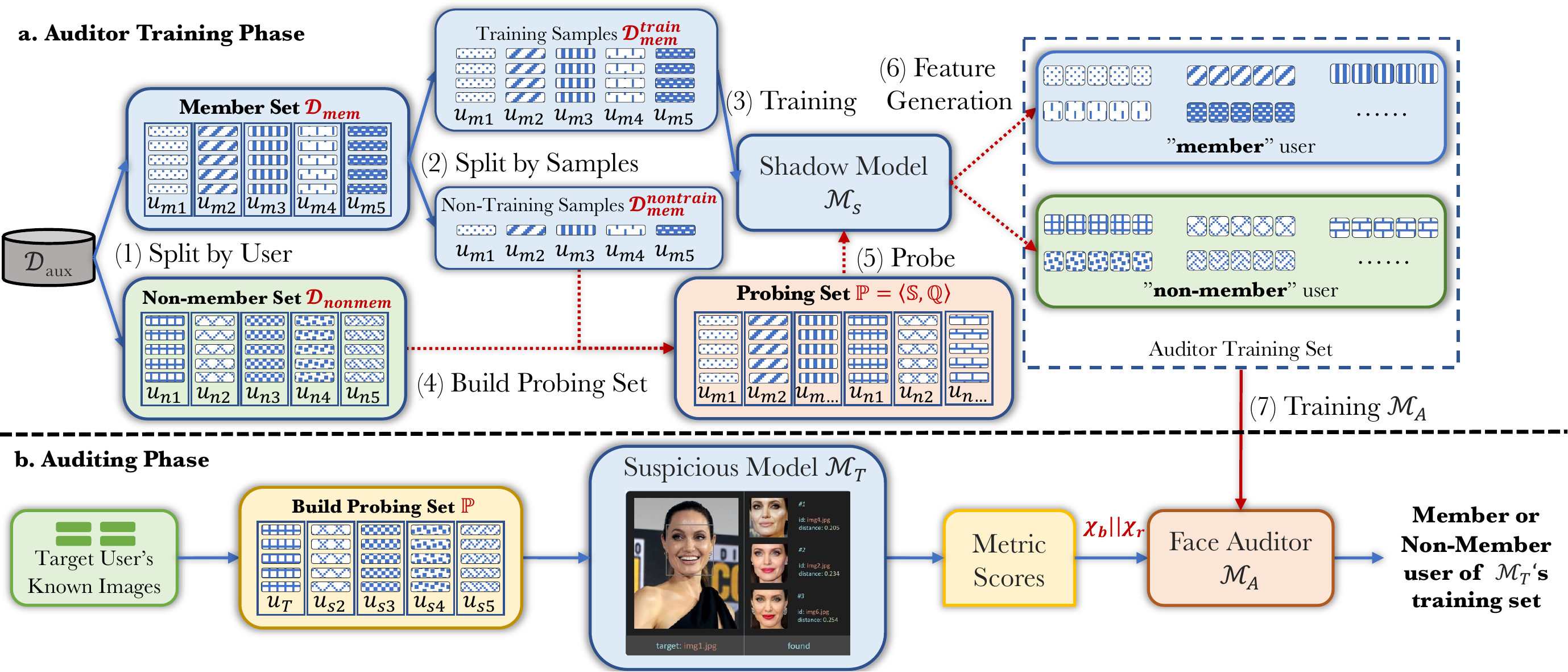}
    \label{subfigure:pipeline}
    \end{subfigure}
    \vspace{-0.5cm}
    \caption{
    Overview of \sysname.  
    There are two phases, training and auditing.  The {\it training phase} is composed of seven steps: 
    (1) The auditor splits its auxiliary dataset into the disjoint member and nonmember set by the user.
    (2) The member set is further split into training and non-training samples. 
    (3) The training samples are used to train a shadow model. 
    (4) The non-training samples and testing set are to form a probing set. 
    (5) We use the probing set to query the shadow model and collect the model outputs. 
    (6) The outputs are labeled by member or non-member, depending on whether the input is from $\dset_{mem}^{nontrain}$ or $\dset_{nonmem}^{}$. 
    (7) We train a supervised binary classifier as our auditing model. 
    In {\it auditing phase}, the auditor builds a probing set with known images from the target users (users to be audited) and then queries a suspicious facial recognition model $\model_{T}$ and collects the corresponding outputs (i.e., similarity scores) as the auditing feature. Feeding these feature vectors into \sysname, the auditor gives a prediction of member or non-member user.}
    \label{figure:pipeline}
\end{figure*}

\mypara{Auditor's Capabilities}
The auditor has a basic knowledge of the facial recognition model, such as metric scores, input format, etc.
To mimic the real-world application, we consider the most challenging setting where the auditor only has {\it black-box} access to the target model.
We assume the auditor can obtain an auxiliary face image dataset.
Note that the auxiliary dataset does not need to contain face images from the same set of users or even the same distribution as the target model; thus, the auditor can utilize some online public datasets to build \sysname which is practical in real-world applications. 
In the auditing phase, the auditor does not need access to the specific face images used to train the target model; instead, it only needs to take a few available face images of the target user.
Furthermore, the auditor can design their own support set (legitimate users) and query set to audit the target model.

\subsection{Overview}
\label{subsection:overview}
\autoref{figure:pipeline} illustrates the overall workflow of \sysname.
Generally, there are two phases, \emph{auditor training} and \emph{target user auditing}.
The auditor training phase aims to train a binary classifier that can distinguish between member users and non-member users.
The general idea is to use the auxiliary dataset $\dset_{aux}$ to train a \textit{shadow model} that mimics the behavior of the target model.
We then design a \textit{probing set} (consisting of support set and query set) to query the shadow model and generate a set of \textit{similarity scores} (between support set and query set), which serves as features to train the auditor model $\model_{A}$.
While most of the existing studies on membership inference are based on the shadow model paradigm~\cite{SSSS17,SZHBFB19,SS19,MXCPZZKX21}, the main challenge lies in constructing the attack/audit features for the attack model. 
For the sample-level membership inference against classification models, the attack features are constructed by feeding the target samples to the target model independently and using the output posteriors as attack features. 
On the other hand, in the user-level few-shot setting, the auditor \emph{does not have the exact images} that are used to train the target model. 
Thus, we need to carefully design a probing set to query the target model and combine the similarity scores as audit features.

In the auditing phase, the auditor collects a set of new face images from the target user and builds a probing set to query the target model.
The auditor then collects the similarity scores returned by the target model as the auditing features and feeds them to $\model_{A}$, which gives a prediction of the member or non-member user.

\subsection{Auditor Training Phase}
\label{subsection:train_phase}
\mypara{Training the Shadow Model}
Assume the auxiliary dataset $\dset_{aux}$ contains face images of $U$ users, and each user has $I$ face images.
We first split $\dset_{aux}$ into two disjoint datasets by users, namely member dataset $\dset_{mem}$ and nonmember dataset $\dset_{nonmem}$.
Recall that, for member users, \sysname does not need to have access to the specific images used to train the auditor model; thus, for the member dataset $\dset_{mem}$, we further split it (by sample) into two disjoint parts, $\dset_{mem}^{train}$ and $\dset_{mem}^{nontrain}$.
We use $\dset_{mem}^{train}$ to train the shadow model and use $\dset_{mem}^{nontrain}$ and $\dset_{nonmem}$to construct the probing set.
To be more clear, all users in $\dset_{mem}$ are the member users, while images in $\dset_{mem}^{train}$ are member samples, and images in $\dset_{mem}^{nontrain}$ are non-member samples.
We follow the procedure described in \autoref{subsection:fsl_models} to construct the support and query set to train the shadow model $\model_{s}$.

\mypara{Constructing the Probing Dataset}
Unlike classical classification models that take a single image as input and output posteriors, few-shot learning models require a support set \support and a query set \query as input and output a sequence of similarity scores (as described in \autoref{subsection:fsl_models}). 
Consequently, generating an auditing feature for few-shot learning is more complex than traditional membership inference attacks against classification models. 
To improve auditing performance, the auditor must carefully design the support set \support and query set \query, rather than directly feeding the training and testing datasets to the shadow model $\model_{s}$ to obtain posteriors. 
For ease of presentation, we call the combination of the support set and query set as probing set $\probe = \langle\support, \query\rangle$.

Since the architecture of \siamese is slightly different from \proto and \relation, we need to design different probing sets for them.

\begin{itemize}[noitemsep]
    \item \mypara{SiameseNet}
    As discussed in \autoref{subsection:fsl_models}, the \siamese model processes the support set separately, which leads to its probing set consisting of a 1-way-$\ell$-shot support set and multiple query images.
    For each probe, we set both the support set and the query images from the same target user (the user to be audited, who may be a member from $\dset_{mem}^{nontrain}$ or a non-member from $\dset_{nonmem}$).

    \item \mypara{ProtoNet \& RelationNet}
    Unlike \siamese, both \proto and \relation takes the support set and query images together as input, forming a $k$-way-$\ell$-shot support set.
    
    We assign the first class of the support set to the target user and select the query images from that user.
    The remaining classes in the support set can be selected from any user, as the similarity scores between these classes and the query images are not used to generate the auditing features. 
\end{itemize}

\mypara{Generating the Auditing Feature}
We use the \textit{similarity scores} between the query image and the support set returned by the shadow model as the \textit{basic auditing feature} $\auditfeat_{b}$.
We use $q$ images in the query set \query, resulting in an auditing feature vector of length $q$.

To further improve the auditing performance, we consider using the image-level similarity between the query image and the support set as additional reference information, referred to as \textit{reference auditing feature} $\auditfeat_{r}$.
In summary, the auditing feature \auditfeat is a concatenation of the basic auditing feature and the reference auditing feature, i.e., $\auditfeat = \auditfeat_{b} || \auditfeat_{r}$.
In this paper, we consider three types of image-level similarity metrics: Directly compare the similarity between image pixels (MSE and CosSim), compare the structural similarity between images (SSIM), and use a deep neural network to compare (LPIPS).
Denote the pixel matrix of two images as $X$ and $Y$, and four metrics can be described as follows.

\begin{itemize}[noitemsep]
    \item \mypara{MSE (Mean Square Error)}
    We first represent the image pair as two pixel vectors $X$ and $Y$, the MSE of these two images is calculated as $\frac{1}{N}\sum_{i=1}^{N}(X_i - Y_i)^2$, where $N$ is the total number of pixels.
    A smaller MSE indicates higher similarity.

    \item \mypara{CosSim (Cosine Similarity)}
    For two pixel vectors $X$ and $Y$, the CosSim is calculated as $\frac{X \cdot Y}{||X||||Y||}$, where~$\cdot$ represents an inner product of two vectors and $||\cdot||$ presents the cardinality of a vector.
    The values of CosSim are in the range of [-1, 1].
    A larger CosSim indicates higher similarity.
    
    \item \mypara{SSIM (Structural Similarity Index Measure)~\cite{WBSS04}}
    It compares two images by considering luminance, contrast, and structure.
    Formally, $SSIM(X, Y) = \ell(X, Y)^\alpha \cdot c(X, Y)^\beta \cdot s(X, Y)^\gamma$, where $\ell(\cdot)$, $c(\cdot)$, $s(\cdot)$ represent luminance, contrast, structure respectively, and $\alpha, \beta, \gamma$ are weight parameters.
    SSIM takes values from the range of [0, 1].
    A larger SSIM value indicates higher similarity. 
    
    \item \mypara{LPIPS (Learned Perceptual Image Patch Similarity)~\cite{ZIESW18}}
    The general idea is to use a pretrained convolutional model to transform the two images $X$ and $Y$ into embeddings, normalize the activations in the channel dimension, and take the $\ell_2$ distance.
    We then average across spatial dimensions and across all layers. 
    A larger LPIPS indicates higher similarity.
\end{itemize}

We conduct empirical experiments in \autoref{subsection:anchor_information} to show that the reference information can effectively improve the auditing performance, and cosine similarity achieves relatively better performance in most settings.

\mypara{Training the Auditing Model}
For all the few-shot learning models, we use $\dset_{train}^{nonmem}$ and $\dset_{test}$ to construct the probing set for member users and non-member users, respectively.
We use a three-layer multi-layer perceptron (MLP) with $100$ hidden neurons as the auditing model.

\subsection{Auditing Phase}
\label{subsection:audit_phase}
To determine whether a target user's face images are used to train the target model, the auditor only needs to take multiple face images from the target user.
Note that these face images are not necessarily used to train the target model.
The auditor then uses the same strategy as the training phase to construct the probing set \probe and generate the auditing feature \auditfeat.
Finally, the auditing feature is fed to the auditing model to determine the membership status of the target user.

\subsection{Discussion}
\label{subsection:technique_novelty}

Here we highlight the technical challenges of \sysname and discuss how we address them in this paper.

\mypara{Mapping Behaviors Differs in Few-shot-based Facial Recognition Models}
In traditional ML models, the inputs and outputs are directly mapped. 
The model outputs either posteriors to known classes or corresponding labels, which enriches the information for a successful membership inference.
However, few-shot-based facial recognition models do not directly map the training data into the corresponding labels (users); instead, they only learn a similarity metric between images.
Even though a class (user) is not seen in the training phase, a facial recognition model could generate an accurate similarity score due to the structural uniqueness of the human face.
Therefore, we cannot determine the membership status of the target users by seeing if the target model can recognize them. 
To solve this challenge, \sysname relies on the fact that images of a member user tend to have higher similarity scores than those of a non-member.
To construct the auditing feature, we concatenate the similarity scores of multiple shots and fix one way in the support set as the target user, which maximizes the similarity of the images from the same user and can be easily implemented in the $k$-way-$\ell$-shot input manner.

\mypara{Black-box Auditing under Domain Shift}
This paper identifies a more practical but strict scenario in which the auditor does not know the underlying training data distribution, which brings more challenges to the shadow model paradigm.
On the one hand, the auditor cannot train a perfect model to mimic the behavior of a targeting facial recognition model.
On the other hand, a black-box auditor can only design the query set to interact with the target model and maximize the difference between member and non-member users.
We trained \sysname slightly differently from the previous shadow model paradigm to solve the domain shift problem.
Concretely, we do not limit our auditor model to the known images but collect unknown images from the training users of the shadow model, increasing the generalization ability of \sysname when disjoint users exist.

\mypara{Well-generalized Models are More Difficult to Audit than the Overfitted Ones} 
A well-generalized model often has a low overfitting level and can behave similarly well on unseen samples from both member and non-member users. 
It is more favorable in practice but more challenging to unify a metric to differentiate between member and non-member users.
As a result, the typical overfitting intuition that guides a successful membership inference attack in the classification model does not apply to few-shot learning settings.
As shown in \autoref{figure:overfitting}, the overfitting level is low in three few-shot facial recognition models.
Thanks to the reference auditing feature, we can build a ground for the anchor images and gradually compare the difference between member and non-member users.
We empirically validate the contribution of reference information in \autoref{subsection:anchor_information}.

\section{Evaluation}
\label{section:evaluation}
In this section, we first describe the experimental setup in \autoref{subsection:experimental_setup} and evaluate the overall auditing performance in \autoref{subsection:end_to_end}.
We then validate the effectiveness of the reference information and investigate the effectiveness of different image-level similarity metrics in \autoref{subsection:anchor_information}.
Finally, we show the transferability of \sysname in \autoref{subsection:transferability}.
We further evaluate the impact of different hyperparameters on the auditing performance in \autoref{app:target_hyperparameters} and investigate the robustness of \sysname when four defense mechanisms are introduced to protect the training images or the target models in \autoref{app:robustness}.

\begin{table*}[th]
    \centering
    \caption{
    Dataset split in detail.
    The dataset was split into two halves for the shadow model and target model, with users being divided equally. 
    We allocated $80\%$ of the users for $\dset_{mem}$ and the remaining $20\%$ for $\dset_{nonmem}$. Within the training set, each user's images were split into two equal parts. One part ($50\%$ as $\dset_{mem}^{train}$) was used to train the shadow/target model, while the other part ($50\%$ as $\dset_{mem}^{nontrain}$) was used to generate the member labels. 
    This split ensured sufficient training data for a well-performed shadow/target model. 
    We keep member and nonmember labels balanced for a fair and accurate evaluation of \sysname.
    }
    \vspace{-0.2cm}
    \label{table:dataset_split}
    \setlength{\tabcolsep}{0.65em}
    \renewcommand{\arraystretch}{1.0}
    \footnotesize
    \begin{tabular}{c c c | c c | c c}
    \toprule
    \multicolumn{3}{c|}{\textbf{Dataset after Prepossessing}}
    & \multicolumn{2}{c|}{\textbf{Target/Shadow Model}}
    & \multicolumn{2}{c}{\textbf{Auditing Model}} \\
    \textbf{Dataset} & \textbf{\#. Users} & \textbf{\#. Images per User} & \textbf{\#. Training Images } & \textbf{\#. Testing Images} & \textbf{\#. Training Images } & \textbf{\#. Testing Images} \\
    \textbf{($\dset$)} & \textbf{$(U)$} & \textbf{$(I)$} & \textbf{$(40\%*U)*(50\%*I)$} & \textbf{$(10\%*U)*(50\%*I)$} & \textbf{$(10\%*U)*(50\%*I)$} & \textbf{$(10\%*U)*(50\%*I)$}\\    
    \toprule
    UMDFaces & 200 & 100 & 4,000& 1,000 & 1,000 & 1,000 \\
    Webface & 827 & 100 & 16,520 & 4,130 & 4,130 & 4,130 \\
    VggFace2 & 5,257 & 100 & 105,140 & 26,285 & 26,285 & 26,285 \\
    CelebA & 6,348 & 20 & 25,392 & 6,348 & 6,348 & 6,348 \\
    \bottomrule
    \end{tabular}
\end{table*}

\begin{table*}[!t]
    \centering
    \caption{
    Target model performance. 
    Higher test accuracy means better representation power and higher overfitting indicates a worse generalization ability of the target model.
    }
    \vspace{-0.2cm}
    \label{table:target_model_performance}
    \setlength{\tabcolsep}{0.14em}
    \renewcommand{\arraystretch}{1.1}
    \footnotesize
    \begin{tabular}{c c | c c c | c c c | c c c | c c c}
    \toprule
    \multicolumn{2}{c|}{\textbf{Dataset}} 
    & \multicolumn{3}{c|}{\textbf{UmdFaces}} & \multicolumn{3}{c|}{\textbf{\textbf{WebFace}}} & \multicolumn{3}{c|}{\textbf{\textbf{VGGFace2}}} & \multicolumn{3}{c}{\textbf{\textbf{CelebA}}} \\
    \midrule
    & \textbf{$\model_{Target}$} & $\model_{\siamese}$ & $\model_{\proto}$ & $\model_{\relation}$ & $\model_{\siamese}$ & $\model_{\proto}$ & $\model_{\relation}$ & $\model_{\siamese}$ & $\model_{\proto}$ & $\model_{\relation}$ & $\model_{\siamese}$ & $\model_{\proto}$ & $\model_{\relation}$ \\
    \toprule
    & \textbf{Train Acc.} & 0.775 & 0.960 & 1.000 & 0.650 & 0.748 & 0.800 & 0.818 & 0.951 & 1.000 & 0.647 & 0.818 & 0.940 \\
    & \textbf{Test Acc.} & 0.500 & 0.794 & \textbf{0.852} & 0.460 & 0.670 & \textbf{0.757} & 0.767 & 0.868 & \textbf{0.943} & 0.603 & 0.802 & \textbf{0.867} \\
    \rowcolor{mygray}
    \cellcolor{white}
    & \textbf{Overfitting} & \textbf{0.275} & 0.166 & 0.148 & \textbf{0.190} & 0.078 & 0.043 & 0.051 & \textbf{0.083} & 0.057 & 0.044 & 0.016 & \textbf{0.073} \\
    \bottomrule
    \end{tabular}
\end{table*}

\subsection{Experimental Setup}
\label{subsection:experimental_setup}

\mypara{Face Datasets}
We perform experiments on four widely used real-world face image datasets: UMDFaces~\cite{BNCRC17}, WebFace~\cite{YLLL14}, VGGFace2~\cite{CSXPZ18}, and CelebA~\cite{LLWT15}. 
The details of the dataset are as follows.

\begin{itemize}[noitemsep]
    \item \mypara{UMDFaces~\cite{BNCRC17}} 
    The original dataset contains 367,888 face images for 8,277 identities. 
    The labels of all the face images are annotated either by human annotators or deep neural networks. 
    The number of samples for each identity varies from 7 to 203.
    \item \mypara{WebFace~\cite{YLLL14}} 
    This is a human face dataset that contains 494,414 face images of 10,575 identities collected from the IMDb website. 
    The number of samples for each identity varies from 2 to 769.
    \item \mypara{VGGFace2~\cite{CSXPZ18}} 
    It is a human face dataset containing 3.31 million images of 9131 identities, 
    which are collected from the Google image search engine.
    These face images show large variations in pose (yaw, pitch, and roll), age, race, lighting, and background.
    The number of samples for each identity varies from 87 to 825.
    \item \mypara{CelebA~\cite{LLWT15}}
    The original dataset contains 202,599 images of 10,177 identities.
    The number of samples for each identity varies from 1 to 35.
\end{itemize}

Since the number of face images for all users is highly unbalanced, to make the experimental results comparable, we filter out the users with a number of images less than 100 (except for CelebA).
For the users having more than 100 images, we randomly sample 100 images for them.
We resize all images to $96\times96$ and evaluate the performance of both the target model and the auditing model. 
Under the setting of \autoref{figure:pipeline}, which indicates half of the images from $40\%$ users are used to train the shadow/target models, we randomly select $10\%$ images from the $40\%$ to generate member labels and $10\%$ testing images for generating non-member labels, they use the data to train and evaluate the performance of \sysname.
We summarize the dataset split in \autoref{table:dataset_split}. 

\mypara{Target Models}
We experiment on three facial recognition system architectures as introduced in \autoref{subsection:fsl_models}, all with the default configurations.
\begin{itemize}[noitemsep]
    \item \mypara{\siamese} 
    Following the setting of~\cite{KZS15}, we implement the \siamese with a four-convolution-layer feature extractor with a ReLU and Max-Pooling for each convolution layer to learn complex patterns in the data. 
    The target model is trained with BCE loss and Adam optimizer.
    \item \mypara{\proto} 
    Following the setting of~\cite{SSZ17}, we implement the \proto with a four-convolution-layer feature extractor with batch normalization and ReLU activation function for each convolution layer.
    The target model is trained with cross-entropy loss and an SGD optimizer with a step scheduler.
    \item \mypara{\relation} 
    Following the setting of~\cite{SYZXTH18}, we implement the \relation with a four-convolution-layer feature extractor and a two-convolution-layer relation network.
    The feature extractor and the \relation are trained with Adam optimizer with a step scheduler. We use MSE loss to train the metric parameters of \relation.
\end{itemize}

\mypara{Metrics}
We use the following four metrics to evaluate the performance of \sysname.
\begin{itemize}[noitemsep]
    \item \mypara{Accuracy}
    We use accuracy to measure the auditing success rate. 
    Concretely, accuracy measures the correctly predicted probing sets to the total probing sets.
    Higher accuracy means better performance.
    \item \mypara{AUC}
    For a binary classification model (our attack model), AUC (the Area Under the Curve) is the measure of the ability of a classifier to distinguish between classes when the decision threshold varies. 
    The higher the AUC, the better the performance of the model at distinguishing between the positive and negative classes. 
    AUC equals 1 indicates perfect prediction, while 0.5 indicates random guessing. 
    \item \mypara{F1 Score}
    F1 Score is a harmonic mean of \textit{precision} (the proportion of true positive cases to the member classes) and \textit{recall} (the proportion of true positive cases to all correctly predicted classes), which can provide a better measure of the incorrectly classified cases than the accuracy metric. 
    A higher F1 Score indicates better auditing performance.
    \item \mypara{False Positive Rate (FPR)} 
    The False Positive Rate (FPR) evaluates the proportion of incorrect ownership claims to the total cases.
    In practice, a higher false positive rate may degrade the credibility of \sysname and cause unnecessary lawsuits.
    In our case, a lower FPR indicates better auditing performance. 
\end{itemize}

\mypara{Experimental Settings}
Following the classical setting of shadow model-based membership inference~\cite{SSSS17,SZHBFB19,SS19,MXCPZZKX21}, we equally split each dataset by users into two disjoint parts, the target set $\dset_{T}$ and auxiliary set $\dset_{aux}$. 
We then split both the target set $\dset_{T}$ and the auxiliary set $\dset_{aux}$ as in \autoref{subsection:train_phase}.
We train the auditing model on $\dset_{aux}$ and evaluate the auditing model on $\dset_{T}$.
We evaluate 5-way-5-shot with 5 queries by default and explore the impacts of different parameters in \autoref{app:target_hyperparameters}.

\begin{figure*}[!ht]
    \centering
    \begin{subfigure}{2\columnwidth}
    \includegraphics[width=\textwidth]{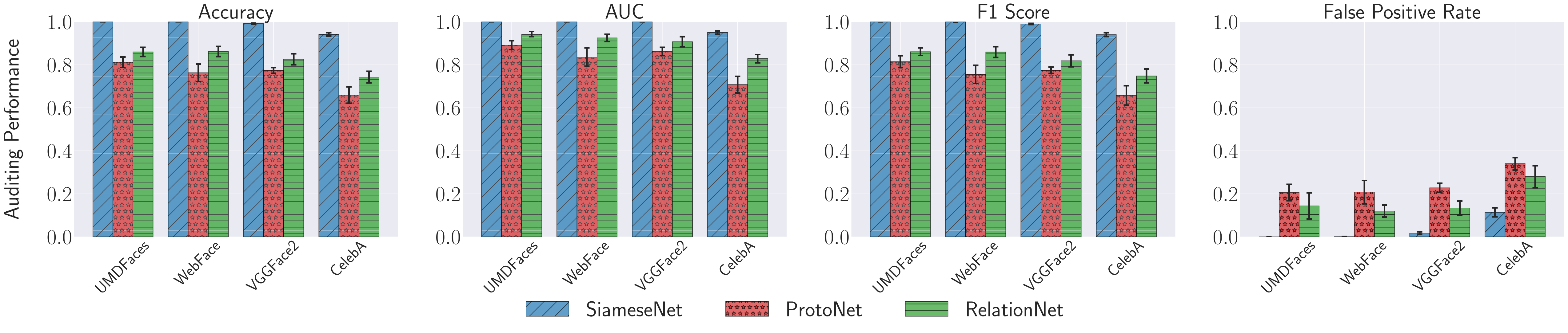}    
    \label{subfigure:class_infer_std}
    \end{subfigure}
    \vspace{-0.8cm}
    \caption{
    Overall auditing performance for four evaluation metrics. 
    We evaluate three model architectures grouped by dataset. 
    We list the auditing performance over four different evaluation metrics in each subfigure.
    }
    \label{figure:class_infer_effectiveness}
\end{figure*}

\mypara{Implementation}
We implement all the target models and the auditing model with Python 3.7 and PyTorch 1.7.
All experiments are run on an NVIDIA DGX-A100 server with 2 TB memory and Ubuntu 18.04 LTS system.
All the experiments are run 10 times with mean and standard deviation reported.

\subsection{Overall Auditing Performance}
\label{subsection:end_to_end}

\mypara{Target Model Performance}
We first investigate the performance of the target models.
\autoref{table:target_model_performance} illustrates the training accuracy, the testing accuracy, and the overfitting (accuracy gap between training and testing datasets) of three target models trained on four face image datasets. 
We first observe that the overfitting level varies across different models but keeps low in most settings.
Besides, the \relation achieves the best testing accuracy, indicating that \relation has the best representation power.

\begin{figure*}[!tpb]
    \centering
    \begin{subfigure}{0.66\columnwidth}
    \includegraphics[width=\textwidth]{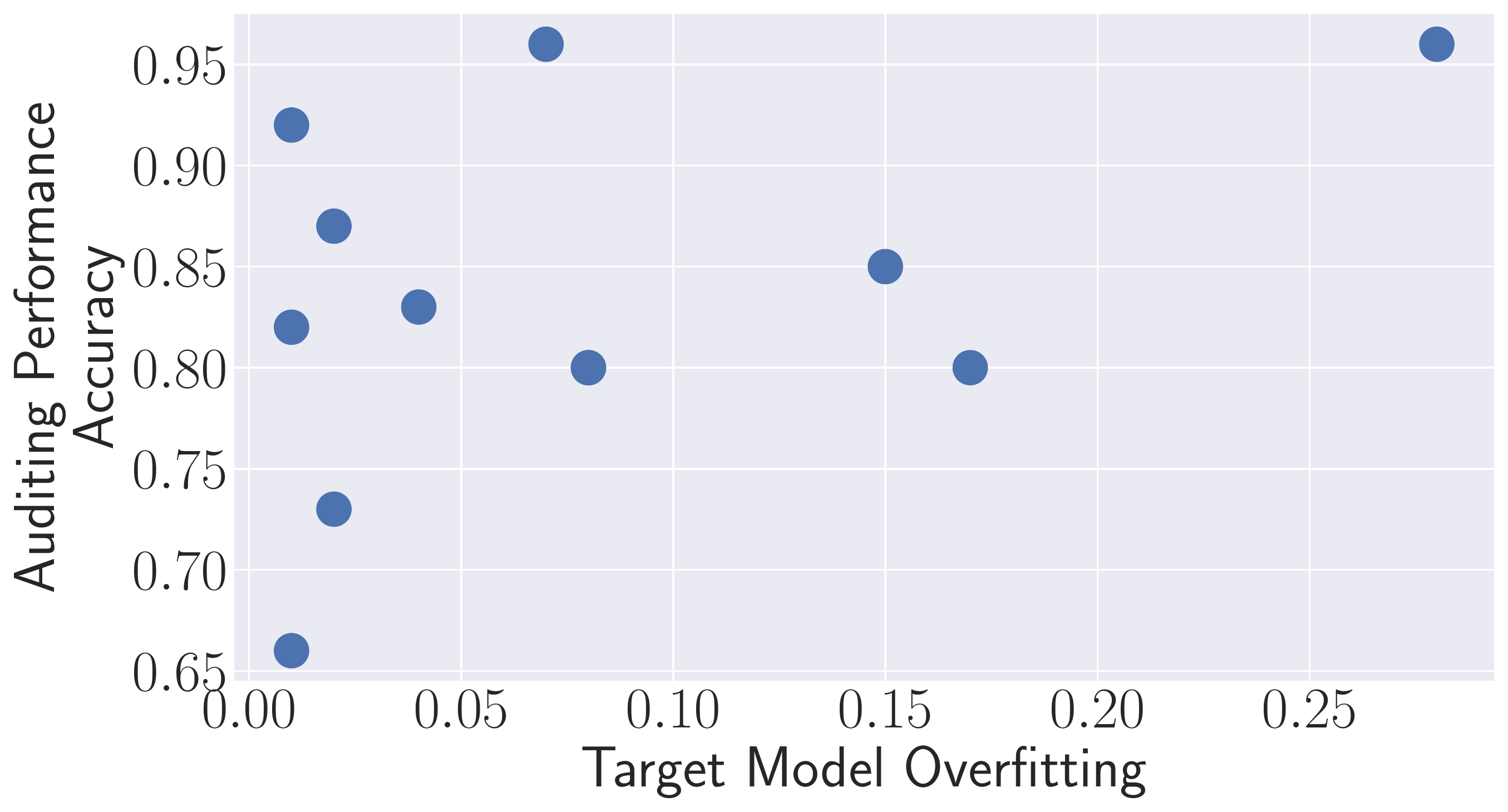}
    \subcaption{Accuracy}
    \label{subfigure:overfitting_accuracy}
    \end{subfigure}
    \begin{subfigure}{0.66\columnwidth}
    \includegraphics[width=\textwidth]{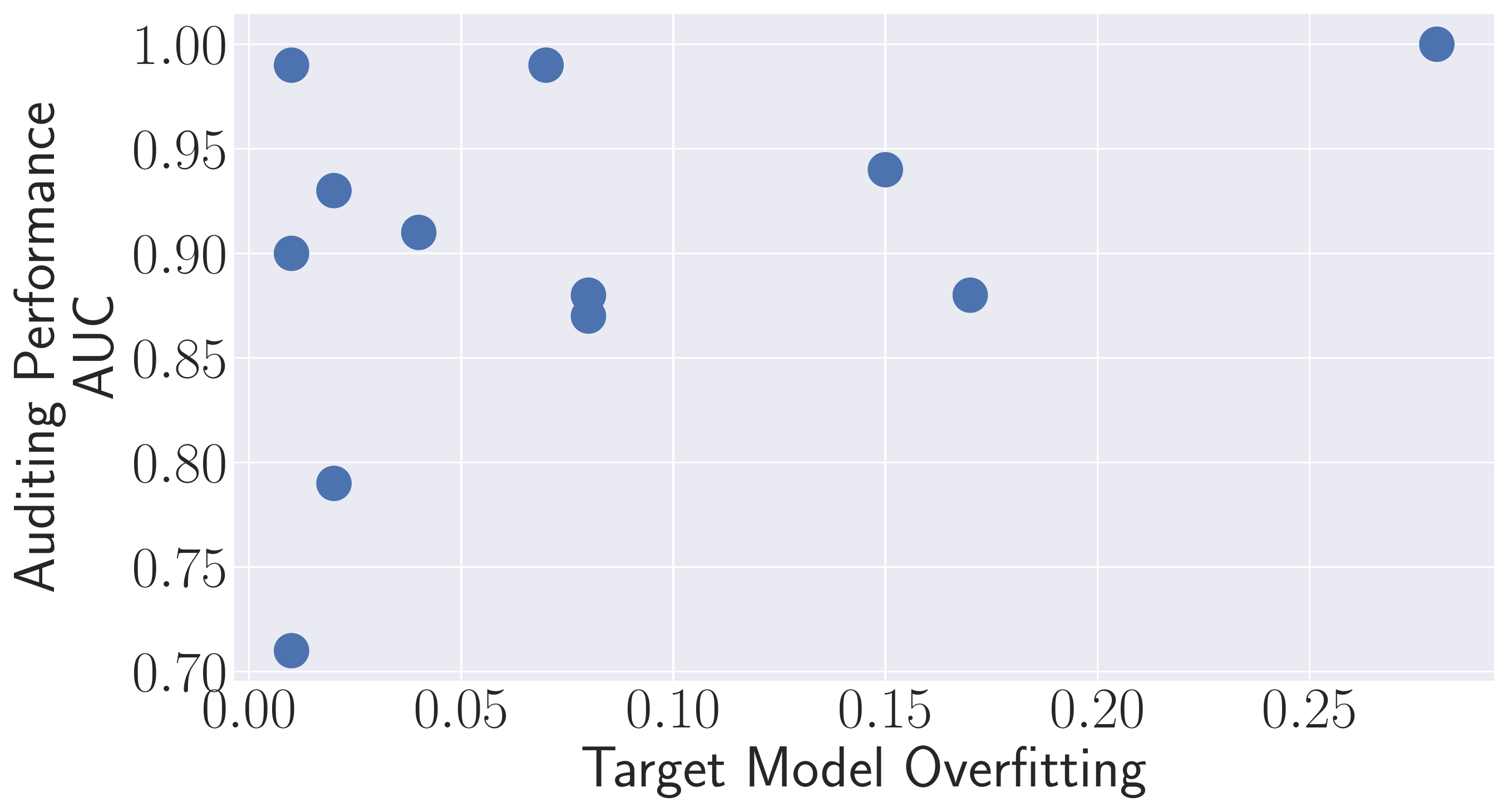}
    \subcaption{AUC}
    \label{subfigure:overfitting_auc}
    \end{subfigure}
    \begin{subfigure}{0.66\columnwidth}
    \includegraphics[width=\textwidth]{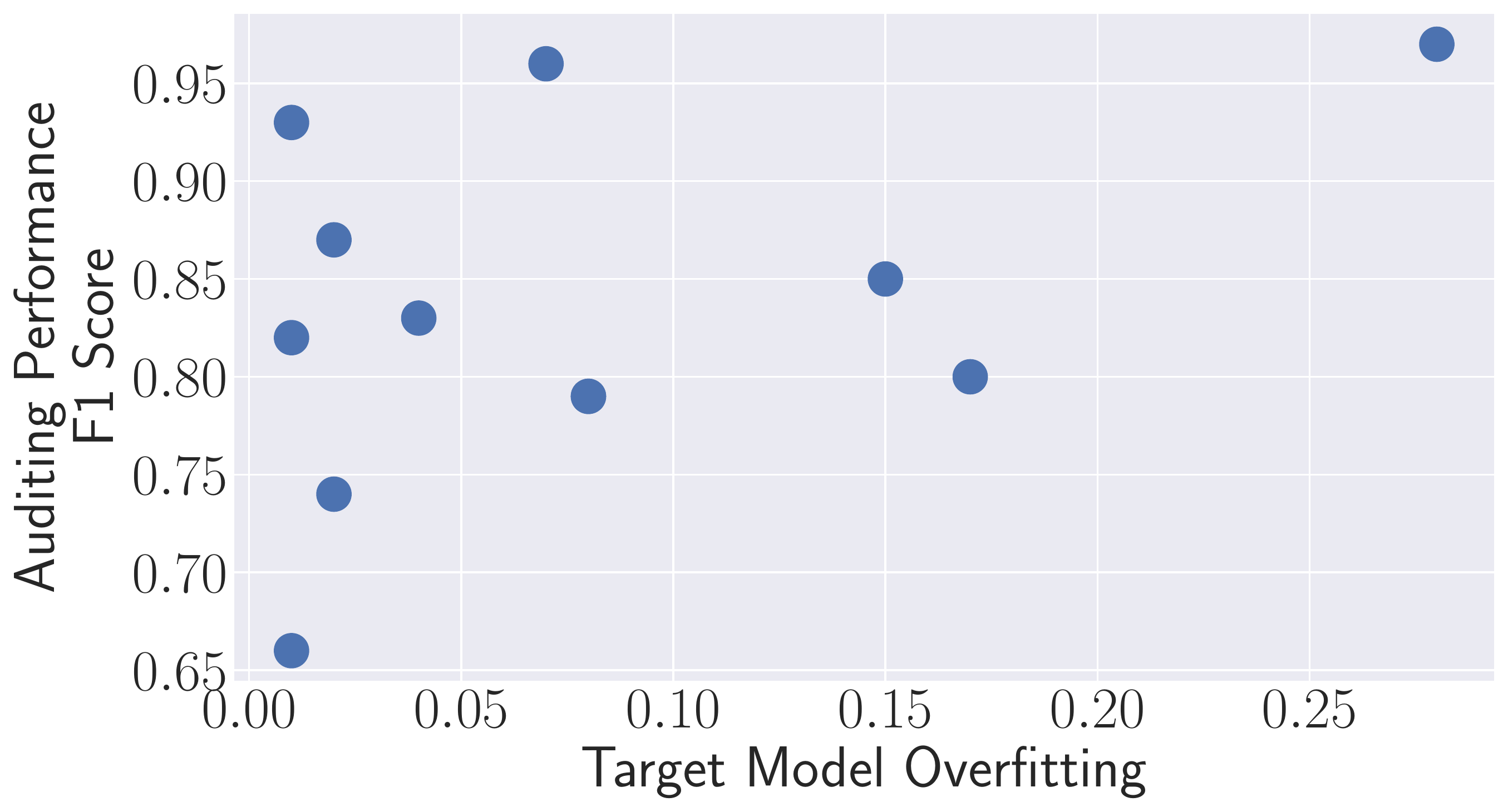}
    \subcaption{F1 Score}
    \label{subfigure:overfitting_f1_score}
    \end{subfigure}    
    \vspace{-0.3cm}
    \caption{
    Relation between target model overfitting and auditing performance. 
    Twelve dots in each subfigure represent the combination of three target model architectures and four datasets.
    The Pearson correlation values between auditing performance (for accuracy, AUC, and F1 Score) and overfitting level are $0.412$, $0.406$, and $0.412$, respectively.   
    }
    \label{figure:overfitting}
\end{figure*}

\begin{figure*}[!t]
    \centering
    \begin{subfigure}{0.66\columnwidth}
    \includegraphics[width=\textwidth]{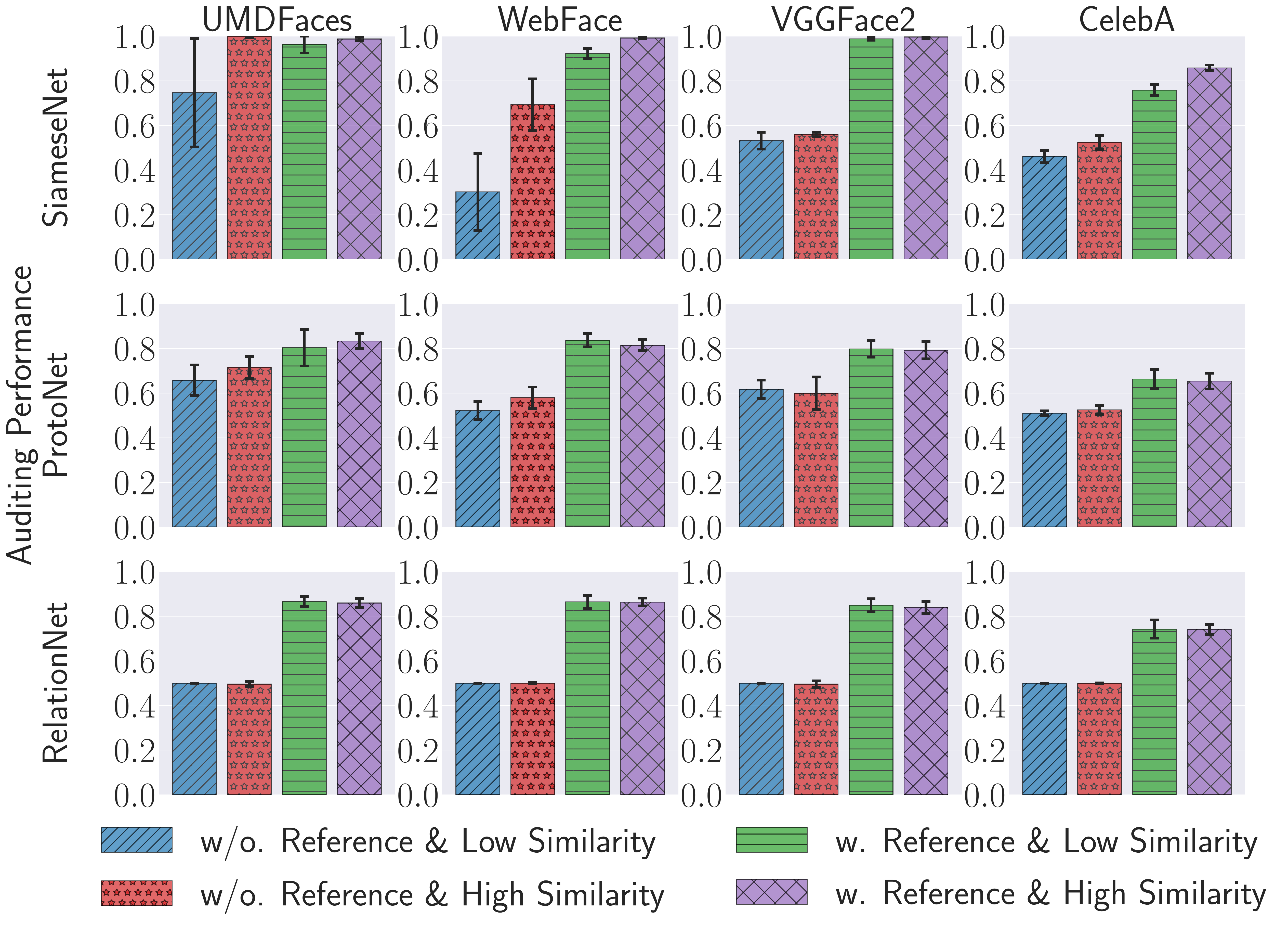}
    \subcaption{Accuracy}
    \label{subfigure:anchor_info_accuracy}
    \end{subfigure}
    \begin{subfigure}{0.66\columnwidth}
    \includegraphics[width=\textwidth]{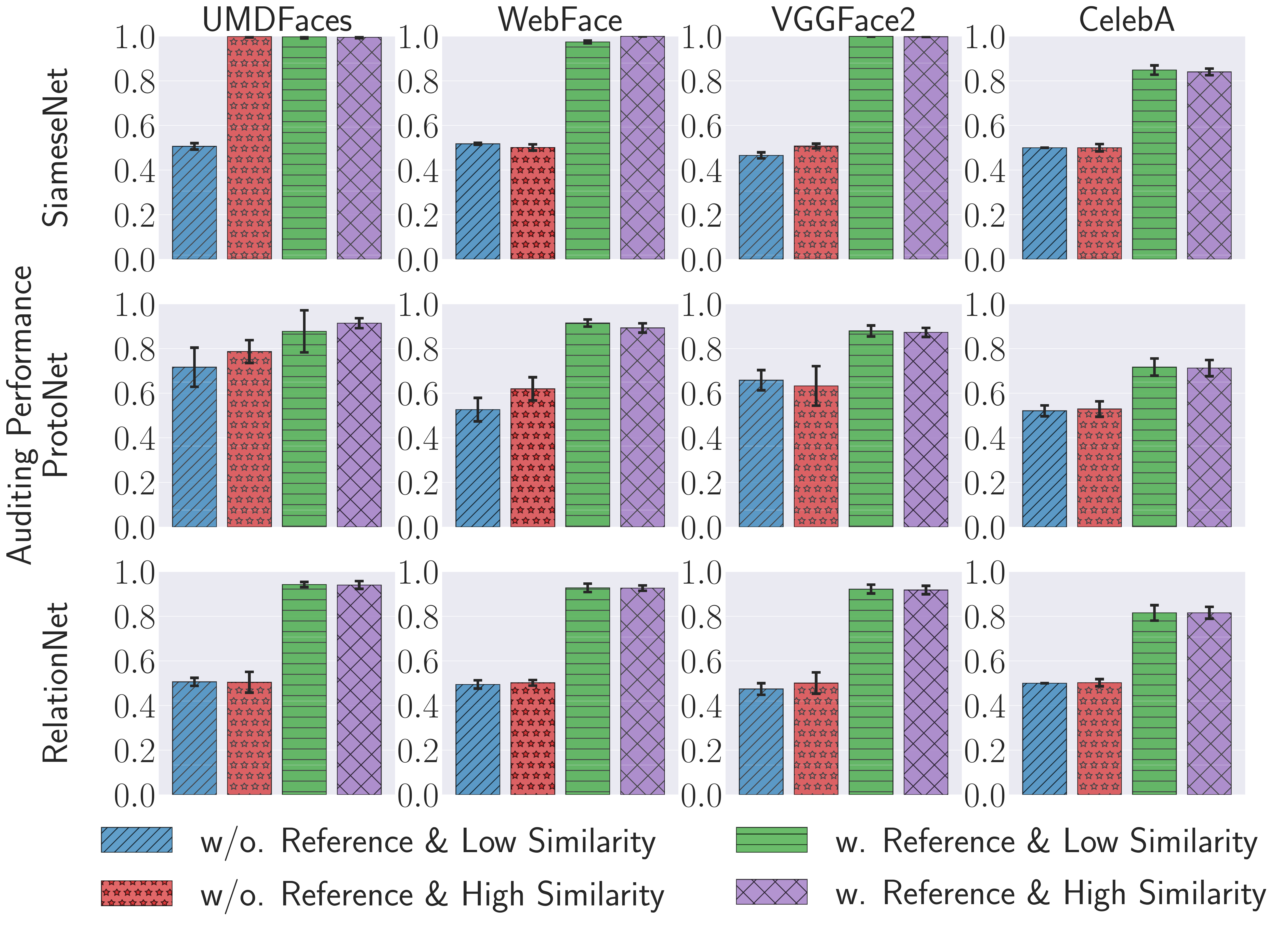}
    \subcaption{AUC}
    \label{subfigure:anchor_info_auc}
    \end{subfigure}    
    \begin{subfigure}{0.66\columnwidth}
    \includegraphics[width=\textwidth]{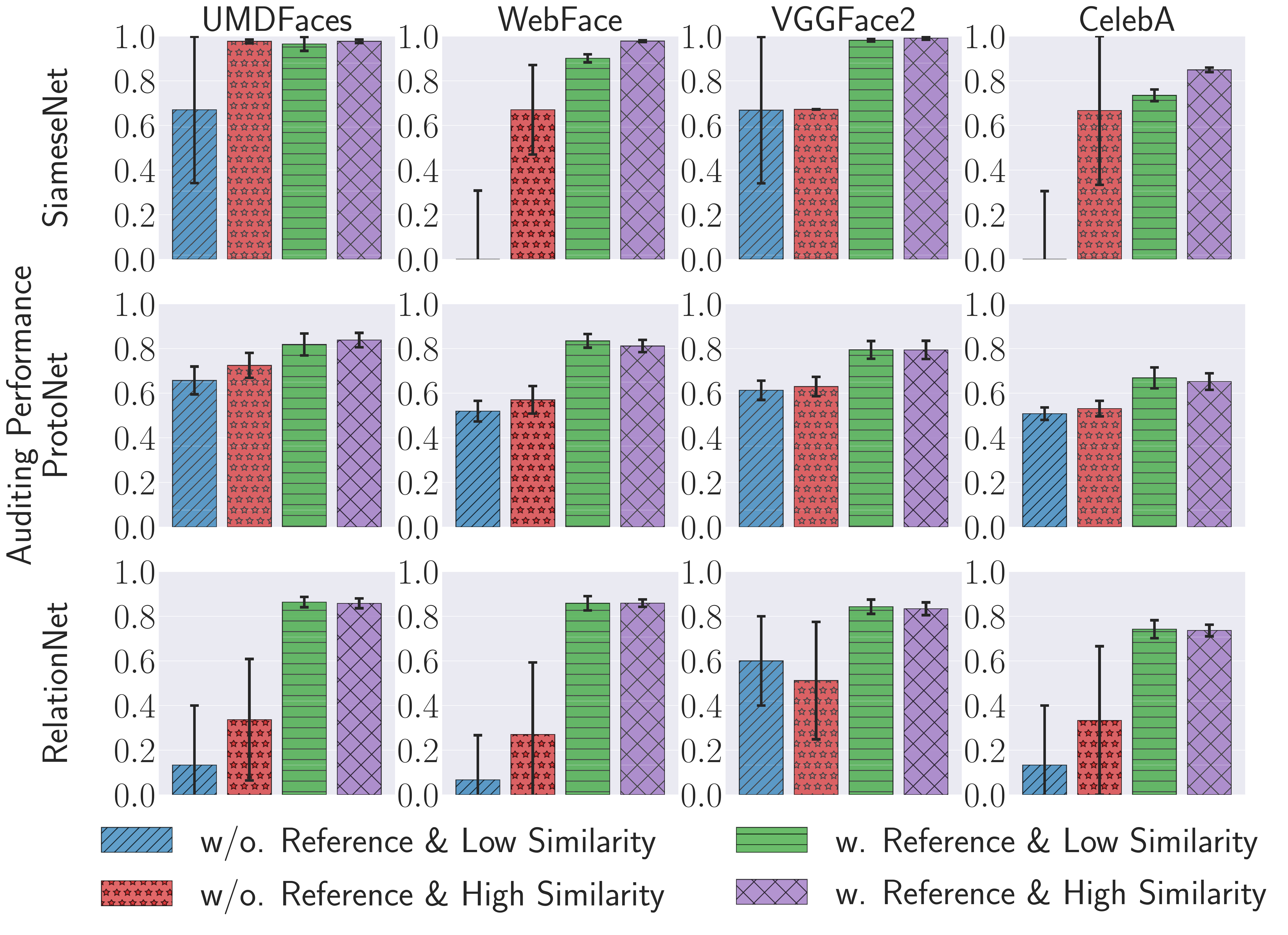}
    \subcaption{F1 Score}
    \label{subfigure:anchor_info_f1_score}
    \end{subfigure} \\
    \vspace{-0.2cm}
    \caption{
    The impact of the reference information and the similarity selection on three metrics. 
    }
    \label{figure:anchor_info}
\end{figure*}

\begin{figure*}[!t]
    \centering
    \begin{subfigure}[c]{0.66\columnwidth}
    \centering
    \includegraphics[width=0.4\textwidth]{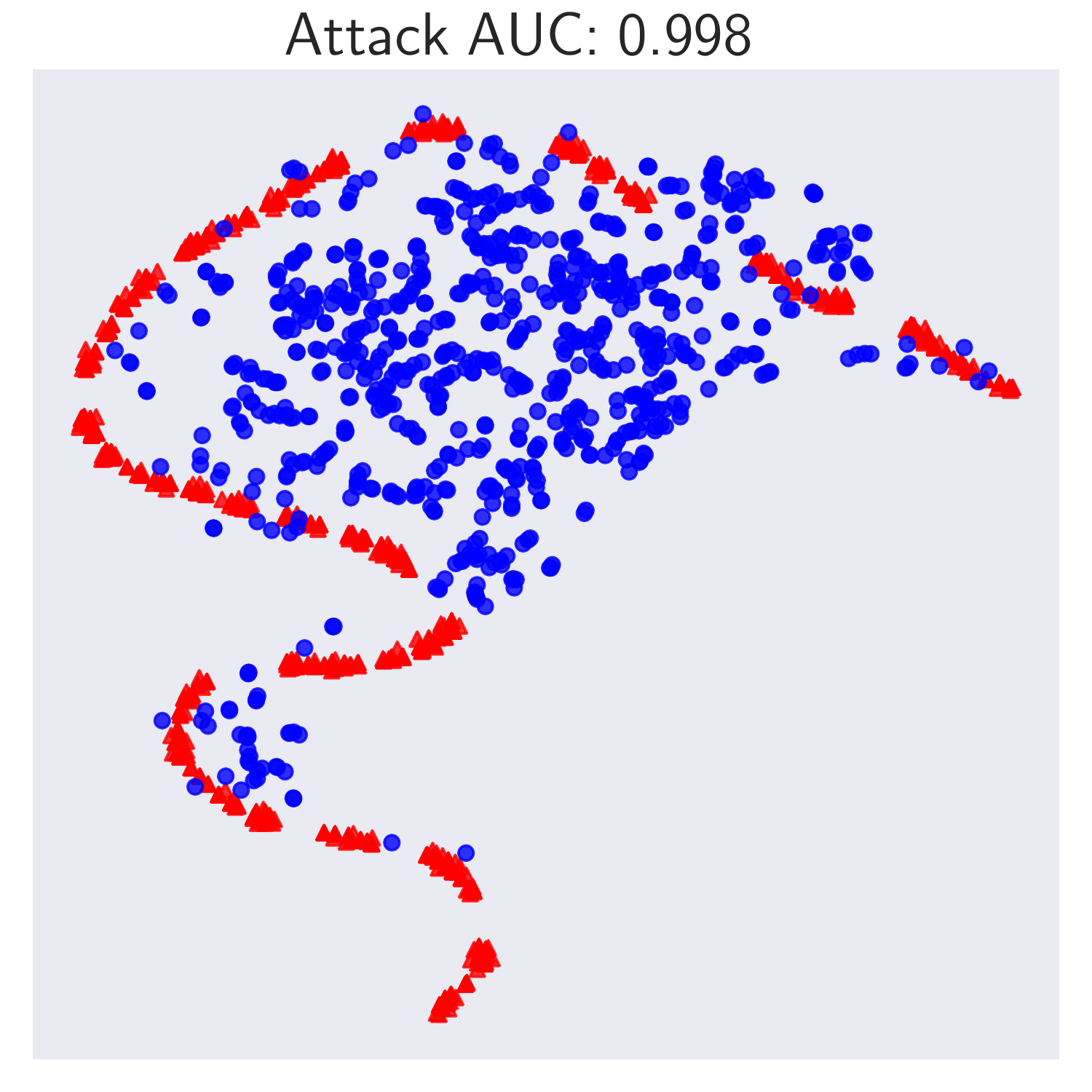}
    \includegraphics[width=0.4\textwidth]{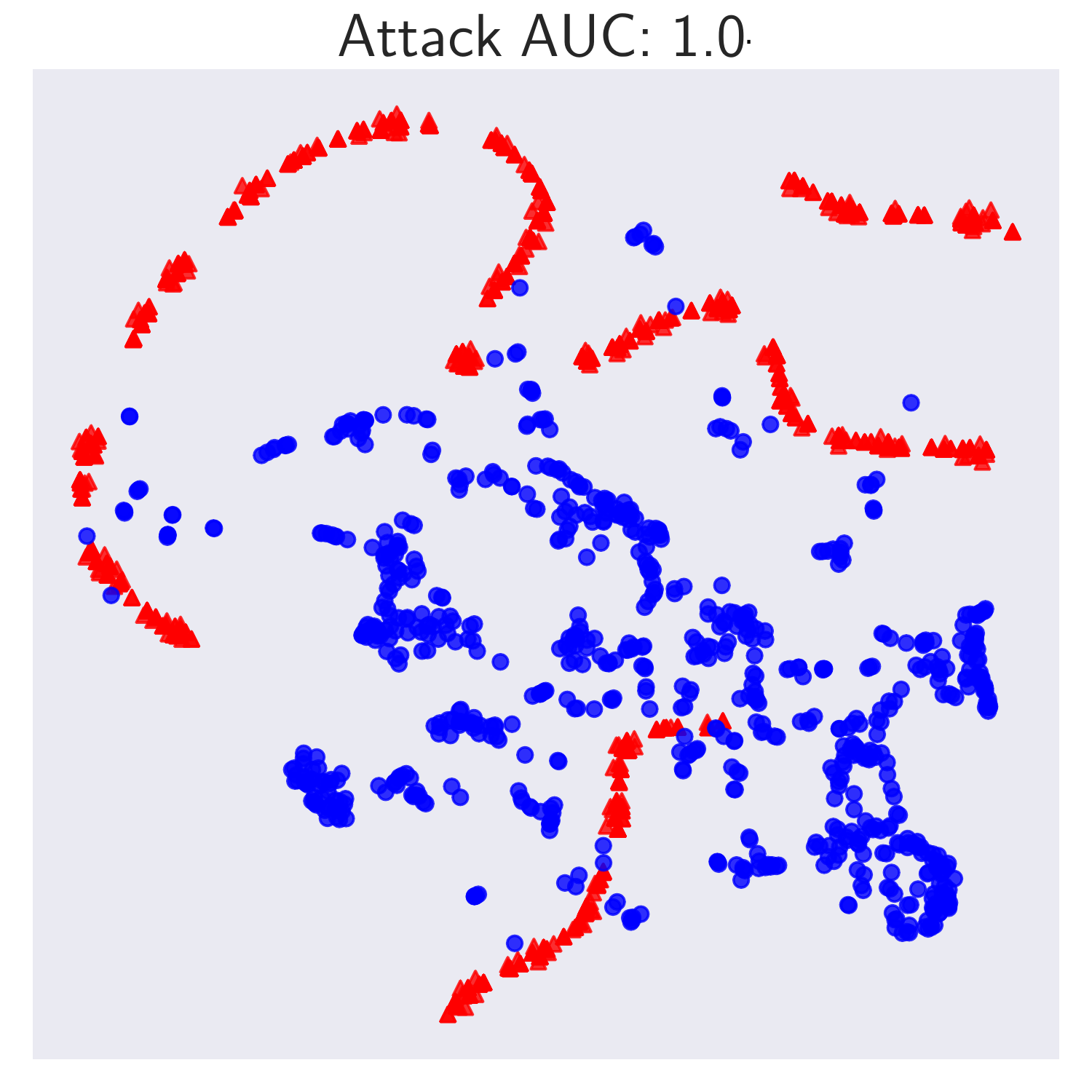}
    \subcaption{\siamese w/o.\& w. reference.}
    \label{subfigure:siamese_tsne}
    \end{subfigure} 
    \begin{subfigure}[c]{0.66\columnwidth}
    \centering
    \includegraphics[width=0.4\textwidth]{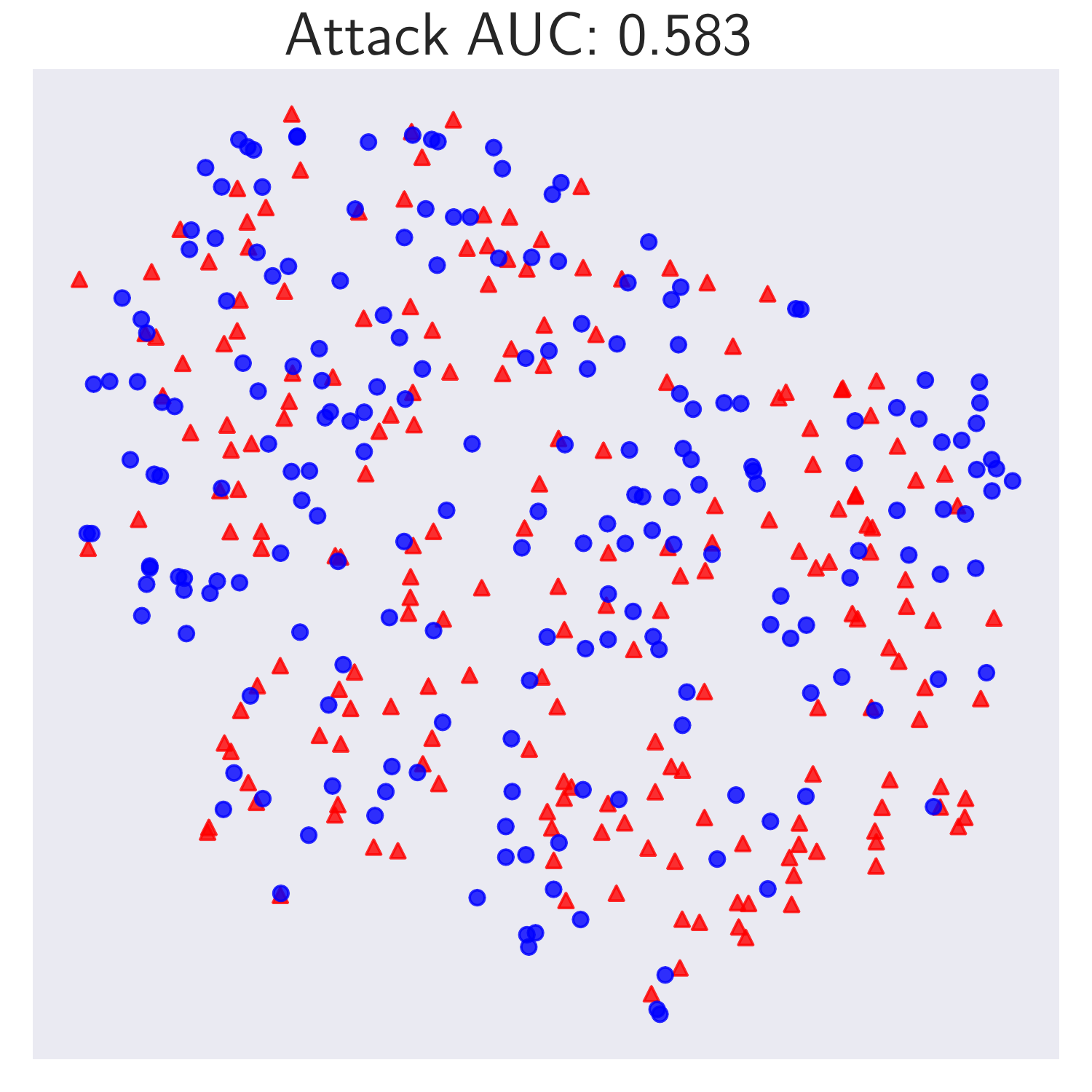}
    \includegraphics[width=0.4\textwidth]{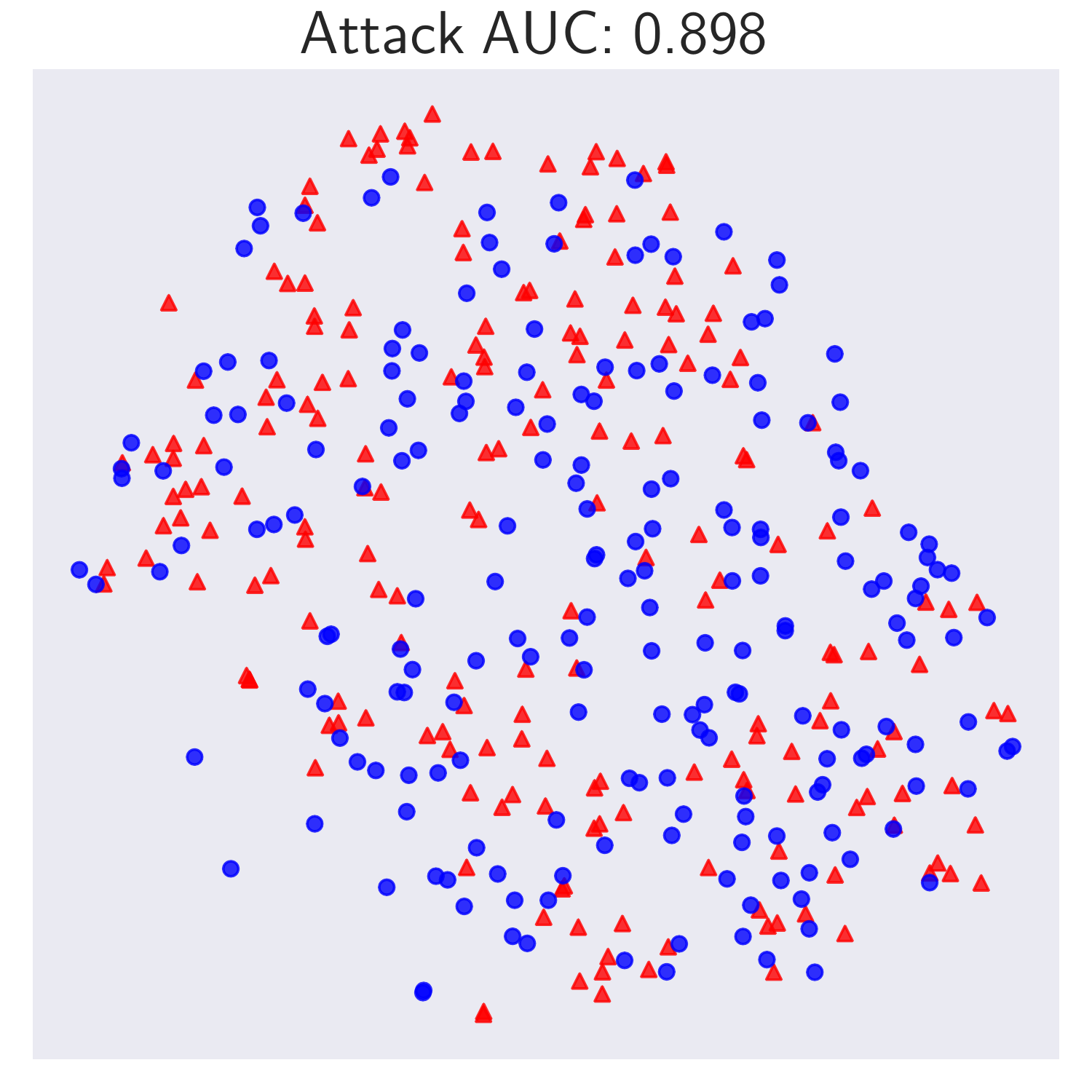}
    \subcaption{\proto w/o.\& w. reference.}
    \label{subfigure:proto_tsne}
    \end{subfigure} 
    \begin{subfigure}[c]{0.66\columnwidth}
    \centering
    \includegraphics[width=0.4\textwidth]{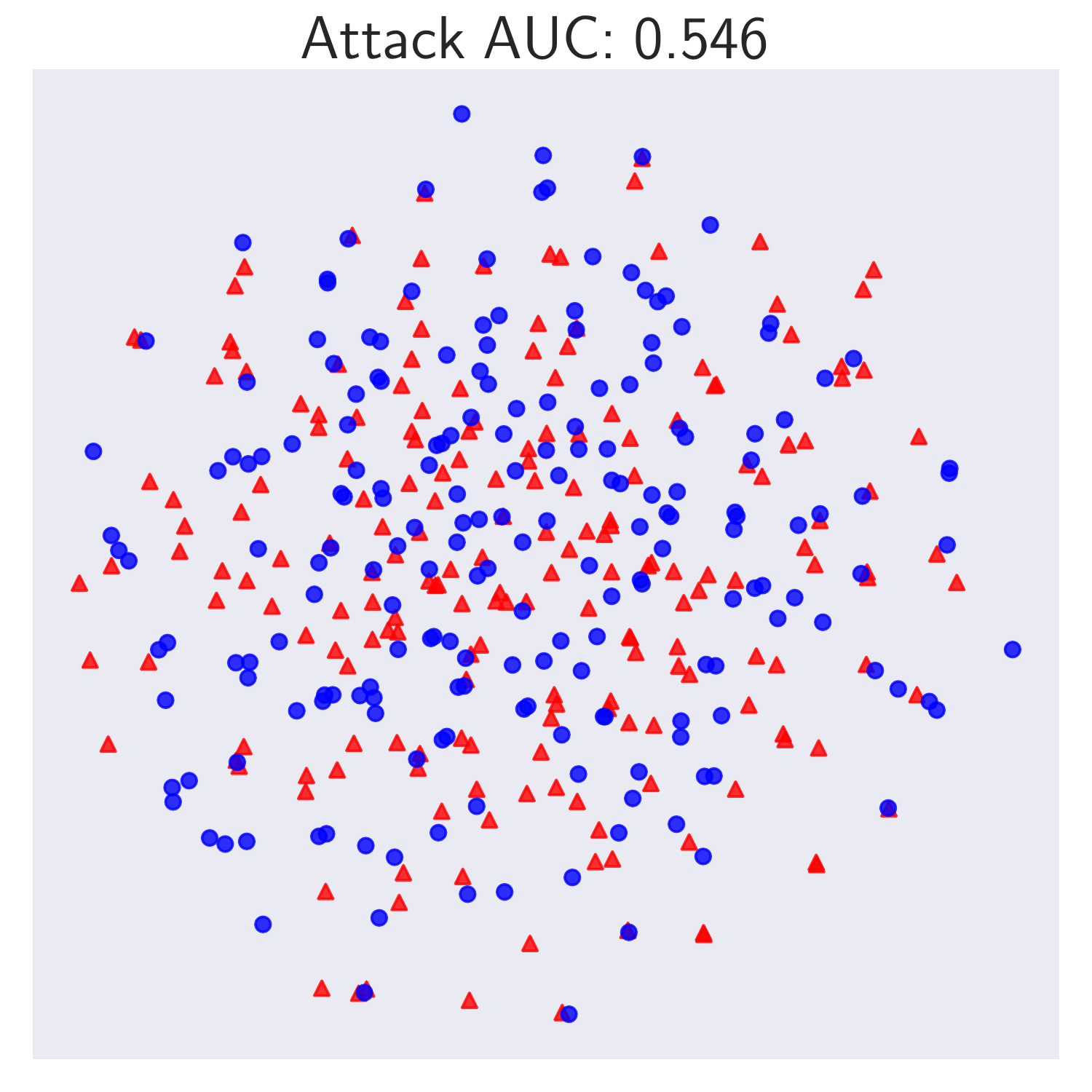}
    \includegraphics[width=0.4\textwidth]{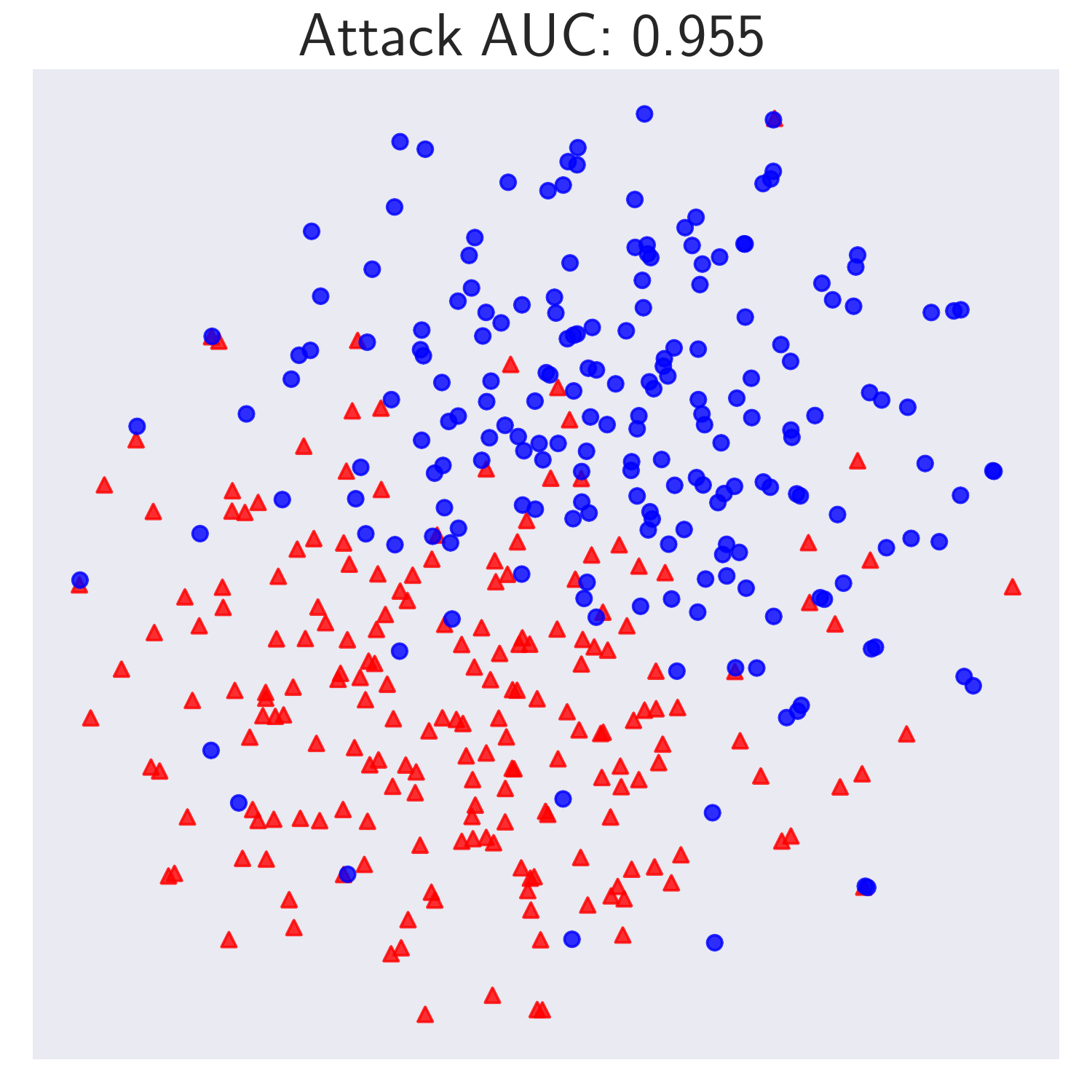}
    \subcaption{\relation w/o.\& w. reference.}
    \label{subfigure:relation_tsne}
    \end{subfigure}
    \vspace{-0.2cm}
    \caption{
    T-SNE visualization on the impact of reference information. Each red triangle is a member sample, and each blue circle is a non-member sample of the UMDFaces dataset. 
    }
    \label{figure:tsne_anchor}
\end{figure*}

\mypara{Auditing Performance}
We then evaluate the overall auditing performance of \sysname.
We conduct experiments on three target models trained on four face image datasets and report the auditing performance with four metrics in \autoref{figure:class_infer_effectiveness}. 

In general, we observe that \sysname achieves good auditing performance for all the target models and datasets.
For instance, \siamese, \proto, and \relation trained on the UMDFaces dataset achieve up to $1.0$, $0.80$, and $0.85$ auditing accuracy, respectively.
We further observe that the auditing performance varies on three different target models.
We achieve the best auditing performance on \siamese and the worst on \proto.
This is due to the different memorization power of the target models.
The memorization power of different models can be explained by the fact that member users' similarity between face images is optimized in the training process. 
\siamese has the highest memorization power since images of the member users are separately optimized in the training process.
In contrast, that of \proto and \relation are optimized together with other classes.
Comparing \proto and \relation, since \relation uses a trainable relation module to compute the similarity scores while \proto directly computes the Euclidean distance; thus \relation has higher memorization power than \proto. 

Comparing different datasets, we observe the best auditing performance on UMDFaces and the worst on CelebA.
This is because UMDFaces has the least users (i.e., 200 users in our experiment), and CelebA has the most users (i.e., 6348 users in our experiment).
Besides, CelebA only contains 20 images for each user; the samples used to represent a user are much fewer than the other three datasets, thus further increasing the challenge for auditing.

\mypara{Remark}
While we observe that \textbf{it is easier to infer a target user's membership status when there are fewer users and each user has more samples in the dataset}.
In practice, it is unnecessary to train few-shot learning-based facial recognition models on face datasets of more than $3k$ users since the objective of the few-shot learning is to learn the similarity information between classes, and $3k$ users are enough for learning a few-shot learning model.

\mypara{Impact of Overfitting}
Overfitting is defined as the model accuracy gap between the training and testing sets, reflecting the learning intensity on training data. 
In general, a model trained with more epochs will result in a higher potential membership leakage risk.
Previous studies have shown that overfitting plays a crucial role in launching a successful membership inference~\cite{YGFJ18,SZHBFB19}.
To investigate the impact of overfitting, we provide a scatter plot showing the relation between the overfitting and the auditing performance in \autoref{figure:overfitting}. 
We observe that higher overfitting indeed leads to better auditing performance. 
Unlike classical sample-level membership inference requiring relatively high overfitting to achieve satisfying inference performance, \textbf{\sysname can achieve good auditing performance even when the overfitting level is low}.
For instance, when the overfitting level is $0.02$, \sysname can achieve $0.93$ auditing accuracy.
On the other hand, the classical sample-level membership inference can only achieve $0.6$ accuracy when the overfitting is $0.02$ (see Figure 2 in~\cite{SZHBFB19}). 

\begin{figure*}[!tpb]
    \centering
    \begin{subfigure}{1.7\columnwidth}
    \centering
    \includegraphics[width=0.8\textwidth]{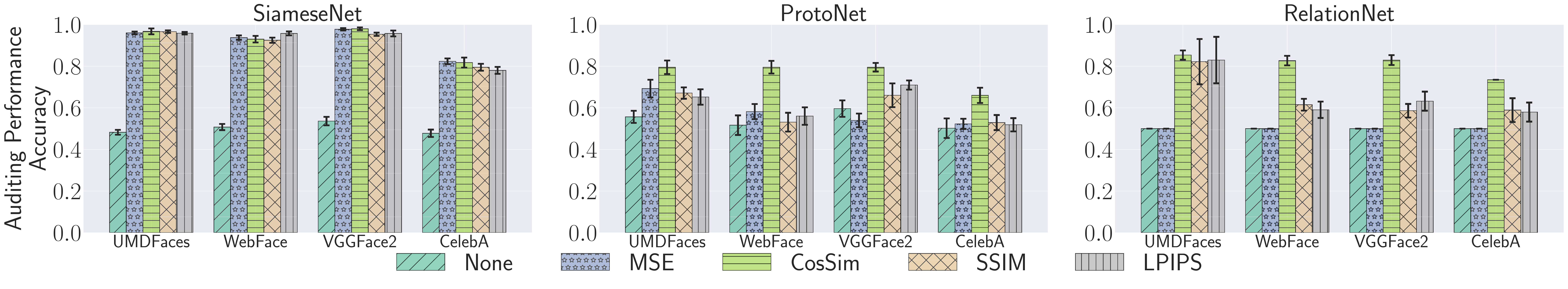}
    \subcaption{Accuracy}
    \label{subfigure:similarity_accuracy}
    \end{subfigure}    
    \begin{subfigure}{1.7\columnwidth}
    \centering
    \includegraphics[width=0.8\textwidth]{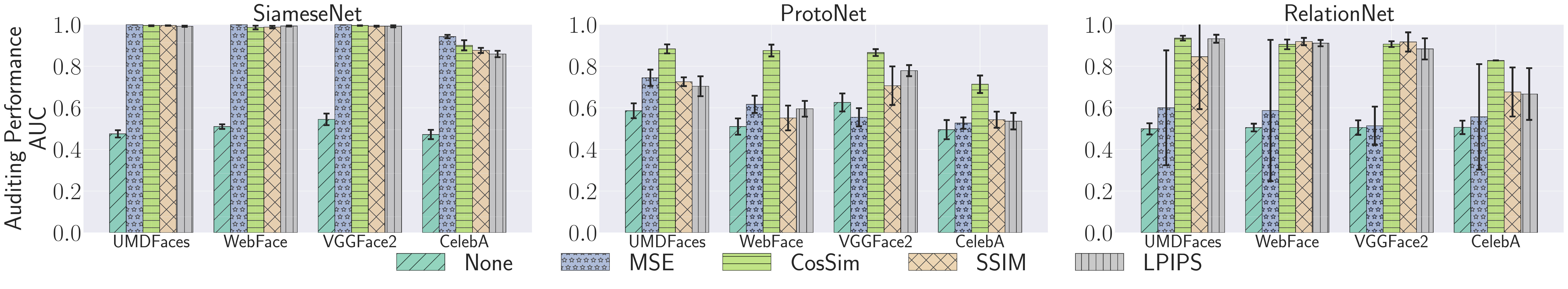} 
    \subcaption{AUC}
    \label{subfigure:similarity_auc}
    \end{subfigure}

    \begin{subfigure}{1.7\columnwidth}
    \centering
    \includegraphics[width=0.8\textwidth]{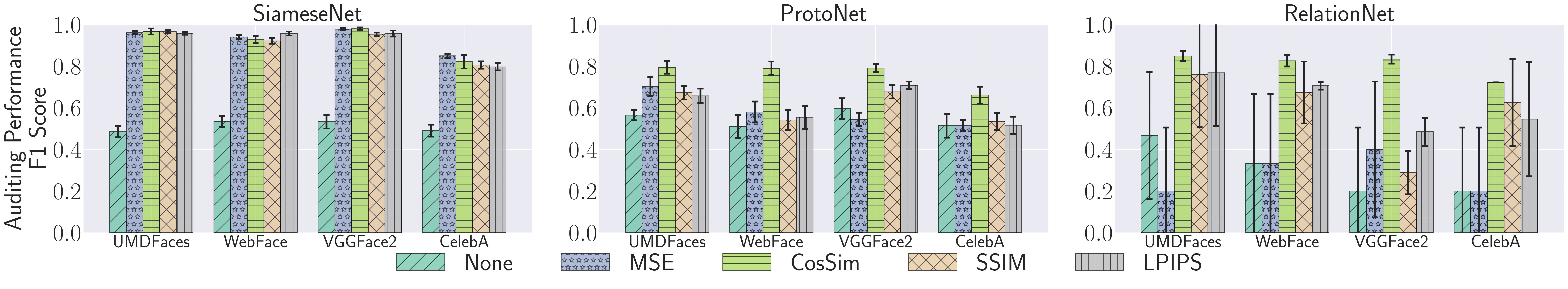}
    \subcaption{F1 Score}
    \label{subfigure:similarity_f1_score}
    \end{subfigure}
    \begin{subfigure}{1.7\columnwidth}
    \centering
    \includegraphics[width=0.8\textwidth]{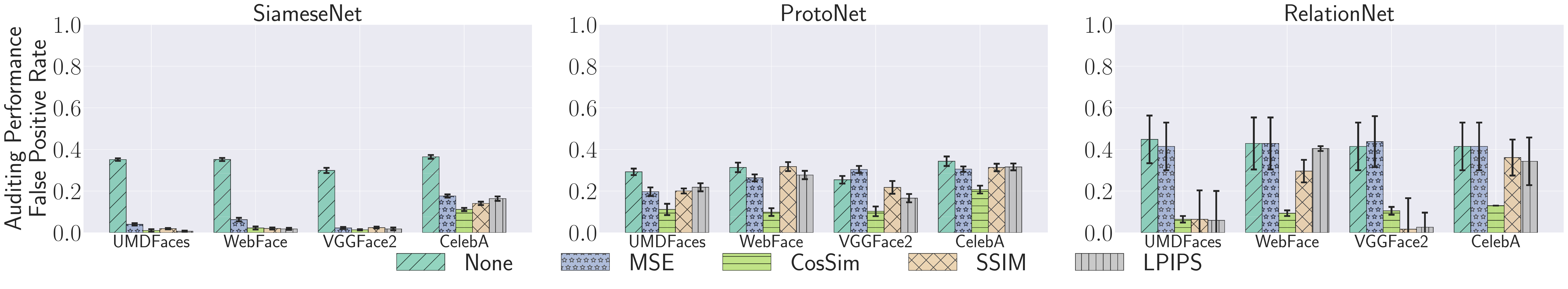}
    \subcaption{False Positive Rate}
    \label{subfigure:similarity_fpr}
    \end{subfigure} 
    \vspace{-0.2cm}
    \caption{
    Auditing performance when using different similarity metrics to generate the reference information. 
    As a supplementary of ~\cite{CZWBZ23}, we report results on four evaluation metrics.
    }
    \label{figure:similarity}
\end{figure*}

\subsection{Effectiveness of Reference Information}
\label{subsection:anchor_information}
As discussed in \autoref{section:audit_method}, the reference information helps to improve auditing performance.
In this subsection, we first validate the effectiveness of the reference information, then investigate the impact of different similarity metrics.

\mypara{Effectiveness}
We conduct experiments on four face image datasets and three target models to validate the effectiveness of the reference information.
The experimental results in \autoref{figure:anchor_info} illustrate that \textbf{exploiting reference information can significantly improve the auditing performance in most of the settings} (by comparing the ``w/o.'' and ``w.'' bars). 

We further explore why the reference information can improve the auditing performance using a t-SNE plot in \autoref{figure:tsne_anchor}. 
Specifically, by comparing \autoref{subfigure:relation_tsne} left and right subfigures, we observe that the member and non-member are much further from each other after exploiting the reference information.
We also observe different effects of reference information on the three target models.
We suspect the reason is that the improvement level by reference information is positively correlated to the memorization power of different models, and \proto has the lowest memorization power in terms of user-level membership inference as discussed in \autoref{subsection:end_to_end}.

\mypara{Impact of Similarity}
Given the reference information of the raw face images, another question is whether to choose query images with high similarity or low similarity to the support set.
To answer this, we compare the auditing performance when choosing the five highest-similarity query images and the five lowest-similarity query images from the testing dataset.
The experimental results are shown in \autoref{figure:anchor_info}. 
By comparing the ``Low Similarity'' and ``High Similarity'' bars, we observe that the original similarity between the query images and the support set only slightly impacts the auditing performance on \proto and \relation.
When auditing the \siamese model, high similarity pairs can enhance the auditing performance.
Take the \siamese trained on the CelebA dataset as an example, the ``Low Similarity'' query images can achieve $0.758$ auditing accuracy, while the ``High Similarity'' query images can achieve $0.858$ auditing accuracy.

In summary, randomly choosing query images from the testing dataset when constructing the probing set can make the auditing model work well in most cases, but \textbf{to achieve the best auditing performance, ``High Similarity'' images are recommended.} 

\mypara{Choice of Similarity Metrics}
We can adopt multiple metrics to measure the similarity between the target image and the support set as discussed in \autoref{subsection:train_phase}. 
\autoref{figure:similarity} illustrates the auditing performance when using different similarity metrics to generate the reference information.
We first observe that all four metrics can achieve relatively high auditing performance on \siamese.
Regarding the auditing performance on \proto and \relation, the performance variance increases among different datasets.
\textbf{In general, CosSim can achieve the best and the most stable performance in most of the settings.}
We posit the reason is that it generates a bounded value (-1 to 1), which tends to be consistent with the normalized input feature of \sysname.

\subsection{Auditor Transferability}
\label{subsection:transferability}
In practice, the auditor might not be aware of the target model's architecture or its training data distribution.
Thus, in this section, we aim to evaluate the transferability of \sysname.
We first evaluate the dataset transferability when the training data of the shadow model comes from a different distribution than the target model, and then evaluate the model transferability when the architecture of the shadow model is different from the target model. As a supplementary of ~\cite{CZWBZ23}, we report results on four evaluation metrics.

\begin{figure*}[!tpb]
    \centering
    \begin{subfigure}{1.8\columnwidth}
    \centering
    \includegraphics[width=0.95\textwidth]{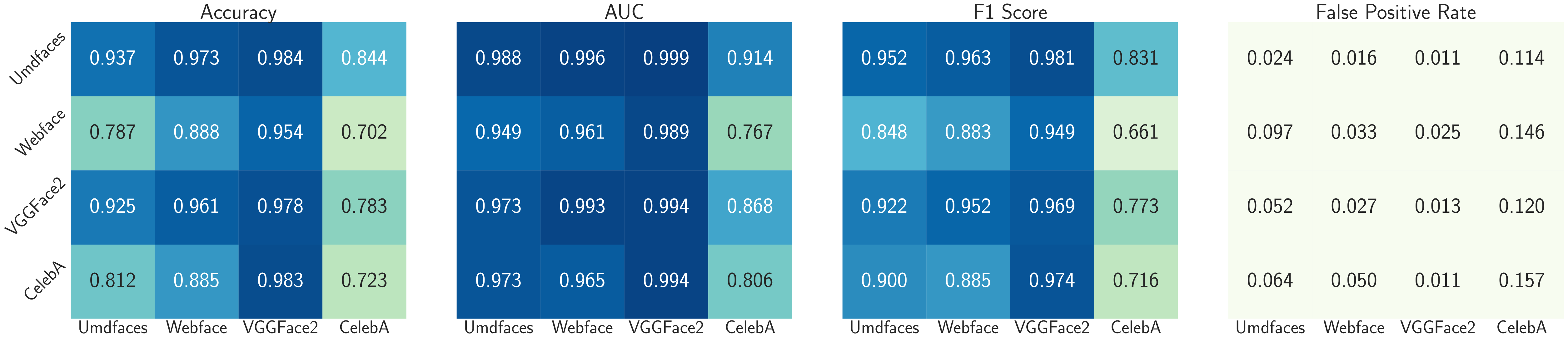}
    \subcaption{\siamese}
    \label{subfigure:siamesenet_dataset_transfer}
    \end{subfigure} \\
    \begin{subfigure}{1.8\columnwidth}
    \centering
    \includegraphics[width=0.95\textwidth]{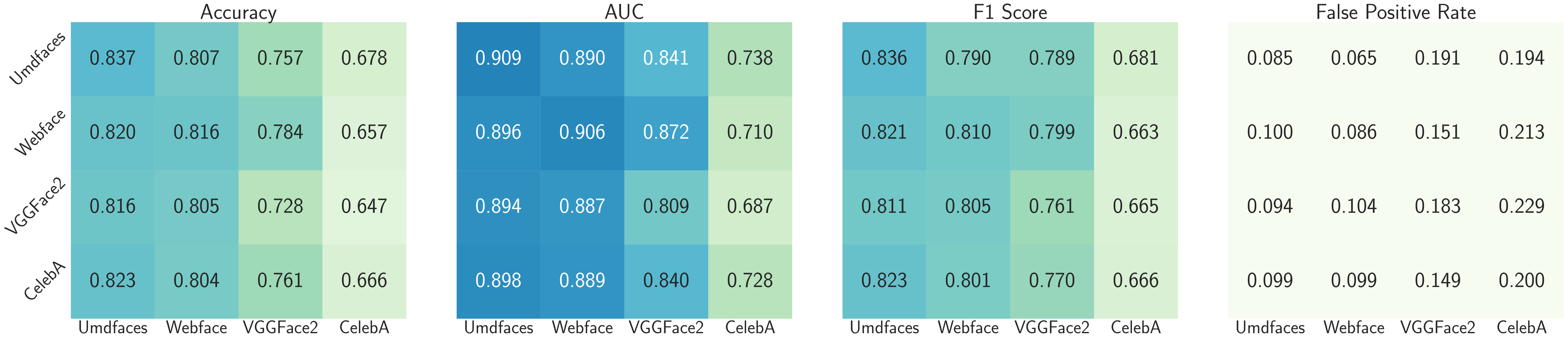}
    \subcaption{\proto}
    \label{subfigure:proto_dataset_transfer}
    \end{subfigure}
    \begin{subfigure}{1.8\columnwidth}
    \centering
    \includegraphics[width=0.95\textwidth]{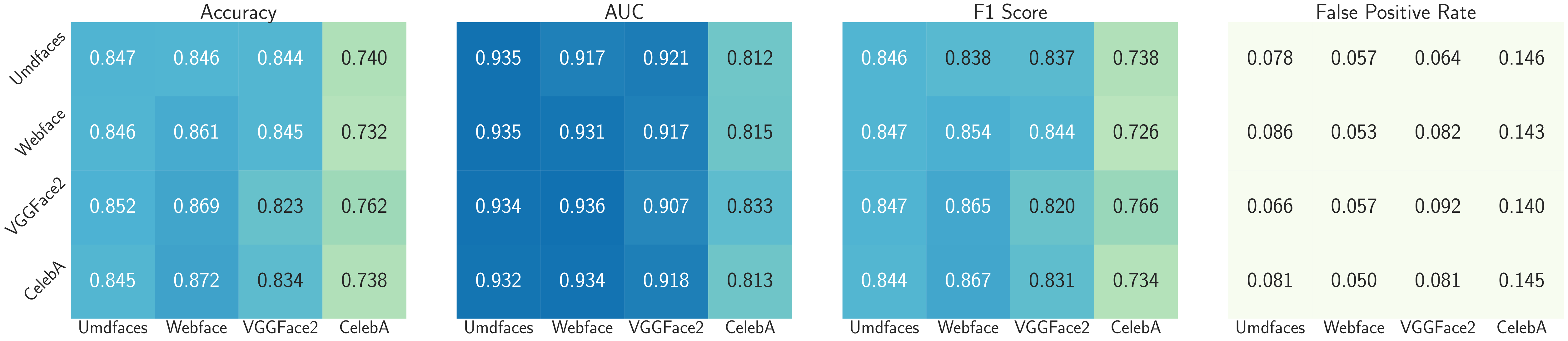}
    \subcaption{\relation}
    \label{subfigure:relation_dataset_transfer}
    \end{subfigure}
    \vspace{-0.2cm}
    \caption{
    Auditing performance under datasets transfer. 
    The x-axis is the dataset used to train the shadow models and probe the target/shadow models.
    The y-axis is the dataset used to train the target models. 
    }
    \label{figure:dataset_transfer}
\end{figure*}

\mypara{Dataset Transferability}
We conduct experiments on three target models.
For each target model, we use one dataset as the auxiliary dataset and the other three datasets as target datasets.
In total, we have 16 combinations.
We report the experimental results for the AUC metric in \autoref{figure:dataset_transfer}. 
In general, we observe that \sysname maintains a good performance when the target dataset and the auxiliary dataset come from different distributions in most of the cases.
For instance, when the auxiliary dataset is VGGFace2, and the target dataset is WebFace, we can achieve up to $0.954$ auditing accuracy, only $0.029$ lower than the same distribution auxiliary dataset.

Two reasons can explain the high auditing performance under dataset transfer settings.
On the one hand, the uniqueness of human faces does not change substantially. 
Once a user's image is seen during the training process of the target model, it is easy to distinguish it from those never seen before. 
Which also shows the severe privacy risks of facial recognition models.
On the other hand, \sysname is trained on a shadow dataset with no user overlap as the target model's training dataset. 
\textbf{The disjoint classes split forces the auditing model to not rely on the overfitting intuition to determine the membership status but learn to discriminate from the metric scores' internal correlations.}

\begin{figure}[!tpb]
    \centering
    \begin{subfigure}{1.0\columnwidth}
    \includegraphics[width=\textwidth]{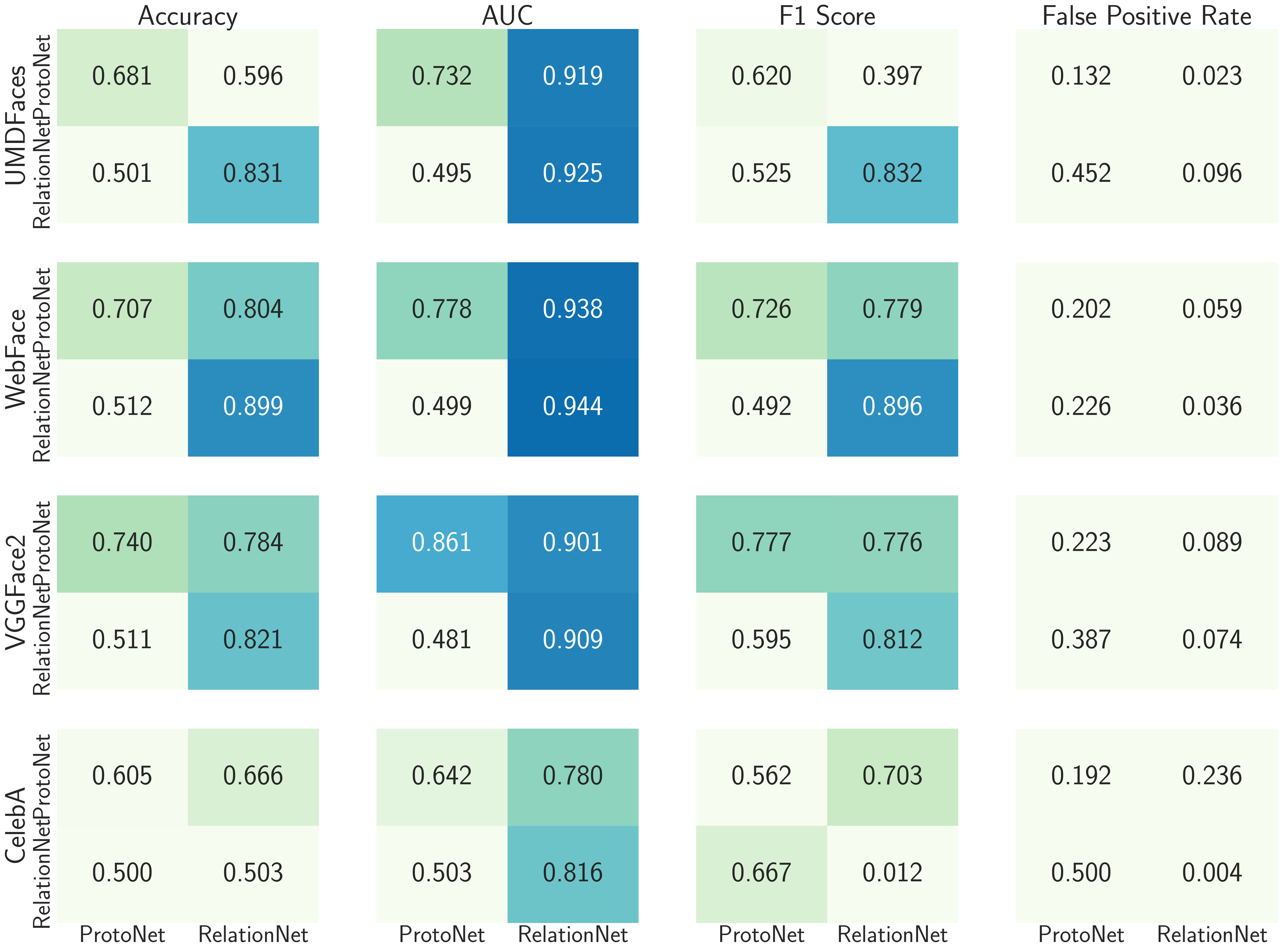}
    \label{subfigure:model_transfer_auc}
    \end{subfigure}
    \vspace{-0.5cm}
    \caption{
    Auditing performance under model transfer.
    In each subfigure, the x-axis represents the target model, and the y-axis represents the shadow model. 
    }
    \label{figure:model_transfer}
\end{figure}

\mypara{Model Transferability}
We conduct experiments between \relation and \proto due to the fact that they share the same input data format and report the experimental results in \autoref{figure:model_transfer}. 
We observe that the auditing performance slightly decreases when the architecture of the shadow model is different from the target model.
The drop is significant when using \relation as the shadow model to audit \proto.
On the contrary, using \proto as the shadow model to audit \relation can achieve better performance.
We suspect the reason is that the linear Euclidean metric of \proto is a particular case of non-linear metric, which supports a pre-trained linear model still work in most cases.

\section{Discussion}
\label{section:discussion}
\mypara{Practical Impacts of \sysname}
\sysname can serve as a complementary tool for existing privacy-protective actions.
Governments and regulators can use \sysname as a tool for enforcing privacy regulations by determining if models are misusing data and violating individuals' privacy rights.
\sysname can be used by individuals as an auditing tool to detect potential misuse of face data.
If a misuse happens, they can take legal actions to correct or withdraw their data (according to GDPR Articles 15, 16, 17, 18). 
\sysname can also be employed by model developers to conduct self-inspection and ensure that their models are compliant with privacy regulations while demonstrating transparency in their data processing practices.

\mypara{Potential Risks of Using \sysname}
While \sysname has the potential to increase transparency for users contributing their data to train a model, it also poses a threat to the intellectual property of the model provider. A malefactor could exploit \sysname to launch user-level membership inference attacks against models with sensitive training data and use \sysname as a stepping stone for other malicious activities, such as attribute inference attacks. 
Knowing that a user is in a sensitive facial recognition-based system's authorized zone could also allow an attacker to design adversarial examples to become an authorized user. 
On a facial recognition-based disease diagnosis system, a known member user might also expose to the privacy of having a particular disease.
Although model developers can introduce protective mechanisms, our evaluation of the robustness of \sysname in \autoref{app:robustness} demonstrates that its inference performance remains high even when subjected to four perturbations. 
In situations where private information is at risk of being inferred by malicious users, criminal laws such as the UK's Data Protection Act 2018 (sections 170 and 171) can deter such activities from occurring~\cite{UKDPA}.

\mypara{Extend to Other Data Domain}
Our paper mainly focuses on facial recognition models, so we use the term ``user-level'' instead of ``class-level''. 
We believe our method can be adapted to other objects as well, and the key challenge lies in choosing the appropriate reference information. For instance, when dealing with text data, a better reference might be the frequency of rare words rather than sentence-level or word-level similarity~\cite{SS19}. 

\section{Conclusion}
\label{section:conclusion}
In this paper, we proposed \sysname to determine whether a target user's face images were used to train a few-shot-based facial recognition system relying on the user-level membership inference.
We carefully designed the probing set to query the few-shot-based facial recognition system.
We further proposed to use the similarity scores between the raw face images as reference information to improve the auditing performance.
We showed that \sysname is robust when the users' face images or the target models are equipped with different defense mechanisms.
In the end, we discuss the practical implications and potential risks of using \sysname.

\section*{Acknowledgments}
We thank all anonymous reviewers and our shepherd for their constructive comments. 
This work is partially funded by the Helmholtz Association within the project ``Trustworthy Federated Data Analytics'' (TFDA) (funding number ZT-I-OO1 4), by the European Health and Digital Executive Agency (HADEA) within the project ``Understanding the individual host response against Hepatitis D Virus to develop a personalized approach for the management of hepatitis D'' (D-Solve) (grant agreement number 101057917), and by NSF grant number 2217071, 2220433 and 2213700.

\bibliographystyle{plain}
\bibliography{clean}
\appendix

\begin{table*}[!t]
    \centering
    \caption{Comparison with Li et al.~\cite{LRL22}. 
    For their method, we use the ($C_{u}, P_{u}$) as the default feature of the auditing model. 
    }
    \vspace{-0.2cm}
    \label{table:embedding}
    \setlength{\tabcolsep}{0.1em}
    \renewcommand{\arraystretch}{1.1}
    \footnotesize
    \begin{tabular}{c c | c c | c c | c c | c c}
        \toprule
        & & \multicolumn{2}{c|}{\textbf{Accuracy}} & \multicolumn{2}{c|}{\textbf{AUC}} & \multicolumn{2}{c|}{\textbf{F1 Score}}  & \multicolumn{2}{c}{\textbf{False Positive Rate}}\\
        \multicolumn{1}{c}{\textbf{Model}} & \multicolumn{1}{c|}{\textbf{Dataset}} & \textbf{Li et al.} & \textbf{\sysname} & \textbf{Li et al.} & \textbf{\sysname} &\textbf{Li et al.} & \textbf{\sysname} & \textbf{Li et al.} & \textbf{\sysname}\\
        \toprule
        & \multirow{1}{*}{UMDFaces} &  65.00 $\pm$ 10.68 & 100.00 $\pm$ 0.00 &  68.35 $\pm$ 3.24 & 100.00 $\pm$ 0.00 & 64.49 $\pm$ 1.18 & 100.00 $\pm$ 0.00 &  35.10 $\pm$ 3.58  & 0.00 $\pm$ 0.00  \\
        & \multirow{1}{*}{Webface} & 63.00 $\pm$ 8.04 & 100.00 $\pm$ 0.00 &  64.47 $\pm$ 3.71  & 100.00 $\pm$ 0.10 & 61.20 $\pm$ 3.56 & 100.00 $\pm$ 0.05 &  38.30 $\pm$ 2.44  & 0.05 $\pm$ 0.10  \\
        & \multirow{1}{*}{VggFace2} & 60.05 $\pm$ 6.32 & 99.17 $\pm$ 0.23 &  63.31 $\pm$ 5.20 & 99.01 $\pm$ 0.51 & 60.61 $\pm$ 7.32 & 97.64 $\pm$ 0.63 & 37.04 $\pm$ 3.97  & 1.75 $\pm$ 0.52  \\
        \multirow{-4}{*}{\rotatebox[origin=c]{90}{\textbf{\siamese}}}
        & \multirow{1}{*}{CelebA} & 57.72 $\pm$ 1.37 & 94.13 $\pm$ 0.81 &   59.85 $\pm$ 4.18 & 95.00 $\pm$ 0.74 & 59.33 $\pm$ 3.81  & 94.01 $\pm$ 0.95 & 44.55 $\pm$ 1.48  & 11.40 $\pm$ 2.13  \\
        \midrule
        & \multirow{1}{*}{UMDFaces} & 50.00 $\pm$ 1.00 & 81.20 $\pm$ 2.40 &  49.92 $\pm$ 6.98 & 89.13 $\pm$ 2.16 & 53.33 $\pm$ 26.67 & 81.47 $\pm$ 3.12 &  80.00 $\pm$ 40.00 & 20.60 $\pm$ 3.78  \\
        & \multirow{1}{*}{Webface} & 50.00 $\pm$ 0.20 & 76.30 $\pm$ 4.08  &  49.33 $\pm$ 3.20  & 83.59 $\pm$ 4.27 & 53.33 $\pm$ 26.67 & 75.54 $\pm$ 4.20 &  80.00 $\pm$ 40.00  & 20.80 $\pm$ 5.42  \\
        & \multirow{1}{*}{VggFace2} & 50.00 $\pm$ 2.40 & 77.40 $\pm$ 1.36 &  48.48 $\pm$ 6.17  & 86.19 $\pm$ 1.90 & 40.00 $\pm$ 
 32.66 & 77.43 $\pm$ 1.48 &   60.00 $\pm$ 48.99 & 22.80 $\pm$ 2.14  \\
        \multirow{-4}{*}{\rotatebox[origin=c]{90}{\textbf{\proto}}}
        & \multirow{1}{*}{CelebA} & 50.00 $\pm$ 0.25 & 65.90 $\pm$ 3.80 &  48.86 $\pm$ 5.51  & 70.77 $\pm$ 3.85 & 53.33 $\pm$ 26.67  & 65.76 $\pm$ 4.53 &  80.00 $\pm$ 40.00 & 34.00 $\pm$ 2.90  \\
        \midrule
        & \multirow{1}{*}{UMDFaces} & 50.20 $\pm$ 0.25 & 86.00 $\pm$ 2.14 & 55.57 $\pm$ 4.78  & 94.28 $\pm$ 1.21 & 44.45 $\pm$ 28.17 & 86.09 $\pm$ 1.75 & 63.40 $\pm$ 45.83  & 14.40 $\pm$ 6.02  \\
        & \multirow{1}{*}{Webface} & 50.00 $\pm$ 0.00 & 86.20 $\pm$ 2.38 & 53.39 $\pm$ 3.48  & 92.53 $\pm$ 1.66 & 48.50 $\pm$ 24.49 & 85.92 $\pm$ 2.53 & 39.60 $\pm$ 48.50  & 12.00 $\pm$ 2.83  \\
        & \multirow{1}{*}{VggFace2} & 49.70 $\pm$ 0.60 & 82.60 $\pm$ 2.58 & 51.68 $\pm$ 4.09  & 90.75 $\pm$ 2.35 & 27.82 $\pm$ 31.74 & 81.85 $\pm$ 2.80 & 41.20 $\pm$ 48.03  & 13.40 $\pm$ 3.20  \\
        \multirow{-4}{*}{\rotatebox[origin=c]{90}{\textbf{\relation}}}
        & \multirow{1}{*}{CelebA} & 49.80 $\pm$ 0.40 & 74.30 $\pm$ 2.71 & 49.16 $\pm$ 2.10  & 82.86 $\pm$ 1.93 & 50.42 $\pm$ 12.49 & 74.79 $\pm$ 3.24 & 86.00 $\pm$ 28.00  & 28.00 $\pm$ 5.10  \\
        \bottomrule
    \end{tabular}
\end{table*}

\section{Comparison with Li et al.\cite{LRL22}}
\label{app:embedding_results}
Recently, Li et al.~\cite{LRL22} propose a user-level membership inference attack against embedding metric models when the adversary can access the embedding of the target model and leverage an auxiliary dataset to train multiple shadow models. 
They use two distance-based features to perform the attack: The average distance of the target user's images to their centroid ($C_u$) and the average pairwise distances of the target user's images ($P_u$).
The authors propose to combine the two \textit{scalar values} as the attack feature.\footnote{The authors do not open-source their code. We thus implement the method of Li et al. by ourselves and conduct experiments on four datasets and three models.} 
We next compare the performance of Li et al. and \sysname.

\mypara{Observation}
\autoref{table:embedding} illustrates the experimental results, and it shows that \sysname outperforms the method of Li et al. in most cases, indicating the features of \sysname are more informative.
Specifically, the method of Li et al. can work on \siamese (0.6-0.7 accuracy).
This is consistent with the results reported in their original paper (note that we use different datasets; thus, the results are slightly different).
However, the performance on \proto and \relation is close to random guess.
We suspect the reason is that \proto and \relation learn the relative distance information between different classes.
The method of Li et al. only considers intra-class correlation but neglects the inter-class correlation. 
While \sysname utilizes the posterior (in \proto and \relation) corresponding to the target user, which inherently considers inter-class correlation. 
Also, with the help of reference information, \sysname can better capture the slight difference between the original image distances and the similarity scores.

\section{Hyperparameter Study}
\label{app:target_hyperparameters}
Recall that we need carefully design a probing set $\probe = \langle\support, \query\rangle$ to achieve the optimal auditing performance. 
We have three important hyperparameters in the probing set i.e., the number of ways $k$, the number of shots $l$, and the number of queries $q$.
We investigate their impacts on auditing performance in \autoref{subsection:way}, \autoref{subsection:shot}, and \autoref{subsection:query}.
As a supplementary of ~\cite{CZWBZ23}, we report results on four evaluation metrics.
Besides, we also investigate the impact of image size in \autoref{subsection:image_size} and the impact of embedding extractor in \autoref{subsection:embedding_extractor}.

\subsection{The Number of Ways $k$}
\label{subsection:way}
\begin{figure*}[!ht]
    \centering
    \begin{subfigure}{2\columnwidth}
    \centering
    \includegraphics[width=0.8\textwidth]{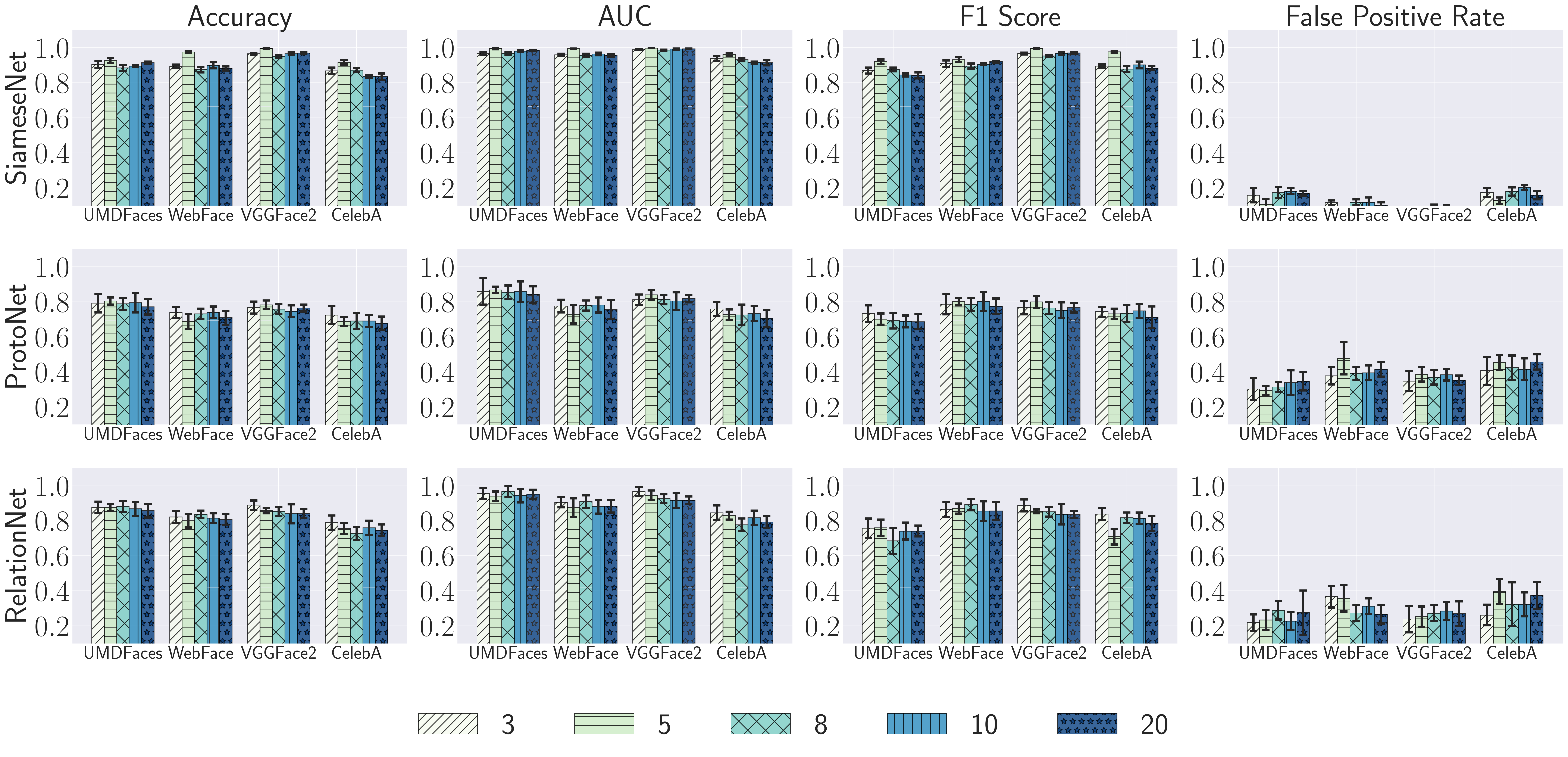}
    \end{subfigure}
    \vspace{-0.2cm}
    \caption{Number of ways ($k$) in the support set.}
    \label{figure:way}    
\end{figure*}

In \autoref{figure:way}, we observe only a slight decrease in the auditing accuracy (less than $4\%$) as we increase the number of ways $k$ in the support set for all three model architectures. 
This parameter affects the search space of the target model (we observe a worse target model performance in more ways of the support set), but it does not significantly affect the auditing model, as we only use the largest similarity score to form the auditing feature.
This also explains why \sysname can work when thousands of users are in the training set of the target model.

\begin{figure*}[!ht]
    \centering
    \begin{subfigure}{2\columnwidth}
    \centering
    \includegraphics[width=0.8\textwidth]{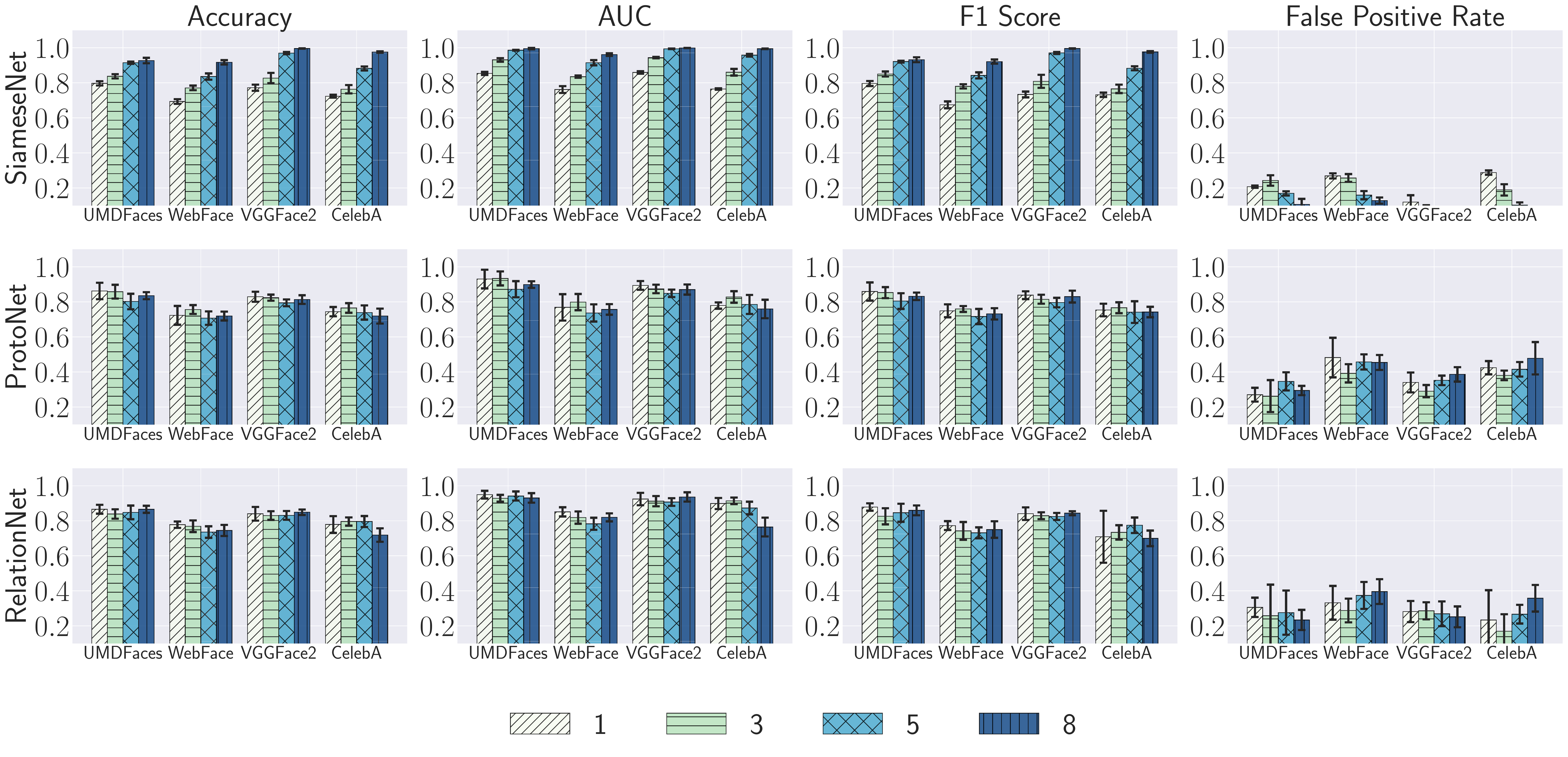}
    \end{subfigure}
    \vspace{-0.2cm}
    \caption{Number of shots ($l$) in the support set.}
    \label{figure:shot}
\end{figure*}

\subsection{The Number of Shots $l$}
\label{subsection:shot}
The results in \autoref{figure:shot} show that increasing the number of shots in the support set leads to a more precise description of a user, as reflected by better target model performance on \proto and \relation. 
However, since \siamese only takes image pairs as input, the target model performance is unaffected by the number of shots.
Interestingly, we found that the auditing performance consistently improved as the number of shots increased for the \siamese. 
We believe this is because \proto and \relation represent each user's multiple images as a whole and calculate inter-class distances to discriminate between multiple users. 
When generating the posteriors, \proto and \relation already consider the influence of multiple shots, resulting in each user being represented as a single vector for comparison, regardless of the number of shots in the support set. 
On the other hand, \siamese takes image pairs per probe, and more shots indicate more diverse probes from a single user. 
This allows for capturing a user's character from multiple probes, leading to an increase in auditing performance.

\subsection{The Number of Query Images $q$}
\label{subsection:query}
\begin{figure*}[!ht]
    \centering
    \begin{subfigure}{2\columnwidth}
    \centering
    \includegraphics[width=0.8\textwidth]{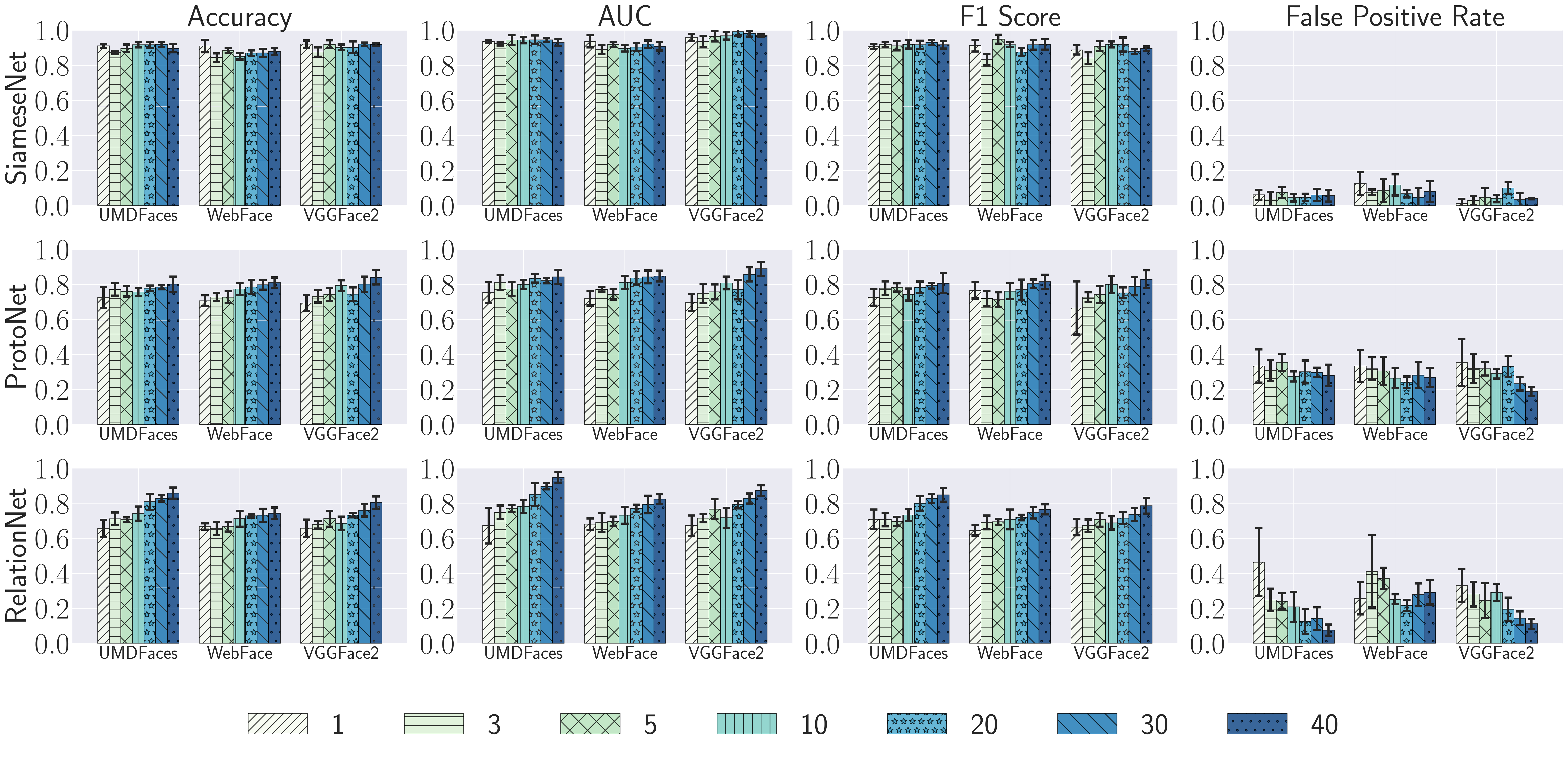} 
    \end{subfigure}
    \vspace{-0.2cm}
    \caption{Number of query images ($q$).}
    \label{figure:query}    
\end{figure*}

We investigated the impact of the number of query images $q$ on three datasets with 100 images per user in their preprocessed dataset, providing a wide range of $q$ values to explore. 
Our results, shown in \autoref{figure:query}, demonstrate that auditing performance improves and the false positive rate decreases as the number of query images increases. 
The rationale is that more query images lead to a broader auditing feature that captures more information about the user, and more images of a user can help distinguish them from other users. 
The increasing trend is more pronounced for \relation and \proto, suggesting that more diverse queries can reveal more information about the underlying training data of the few-shot facial recognition models, especially when the model has a higher memorization ability. 

\subsection{The Image Size}
\label{subsection:image_size}
By default, we resize all the face images to $96\times96$. 
In practice, image quality is affected by size. 
We resize images into $32\times32$, $96\times96$, and $112\times112$, respectively, and show the auditing performance in \autoref{table:image_size}. 
The experimental results show that increasing the image size helps to improve the auditing performance in most cases.
For instance, the auditing accuracy on \siamese with image size $32\times32$ is $0.781$. After we resize all images to $112\times112$, the auditing accuracy increases to $0.858$.

\begin{table*}[t]
    \centering
    \caption{Auditing performance (measured by AUC) under different image sizes.}
    \vspace{-0.2cm}
    \label{table:image_size}
    \setlength{\tabcolsep}{0.2em}
    \renewcommand{\arraystretch}{1.1}
    \footnotesize
    \begin{tabular}{c |c c c |c c c |c c c }
    \toprule
    \multicolumn{1}{c|}{\textbf{Target Model}} 
    & \multicolumn{3}{c|}{\textbf{\siamese}}
    & \multicolumn{3}{c|}{\textbf{\proto}} 
    & \multicolumn{3}{c}{\textbf{\textbf{\relation}}} \\
    \midrule
    \textbf{Dataset} & \textbf{$32\times32$} & \textbf{$96\times96$} & \textbf{$112\times112$} & \textbf{$32\times32$} & \textbf{$96\times96$} & \textbf{$112\times112$}  & \textbf{$32\times32$} & \textbf{$96\times96$} & \textbf{$112\times112$} \\
    \toprule
    UMDFaces & 0.976 $\pm$ 0.007 & \textbf{0.996 $\pm$ 0.003} & 0.995 $\pm$ 0.002 & 0.836 $\pm$ 0.019 & \textbf{0.883 $\pm$} 0.022 & 0.853 $\pm$ 0.041 & 0.891 $\pm$ 0.025 & \textbf{0.935 $\pm$ 0.012} & 0.900 $\pm$ 0.021 \\
    WebFace  & 0.966 $\pm$ 0.009 & \textbf{0.986 $\pm$ 0.009} & 0.956 $\pm$ 0.015 & 0.873 $\pm$ 0.027 & 0.875 $\pm$ 0.028 & \textbf{0.876 $\pm$ 0.023} & 0.905 $\pm$ 0.012 & \textbf{0.905 $\pm$ 0.023} & 0.900 $\pm$ 0.022 \\
    VGGFace2 & 0.998 $\pm$ 0.003 & 0.996 $\pm$ 0.002 & \textbf{0.998 $\pm$ 0.002} & 0.862 $\pm$ 0.027 & 0.866 $\pm$ 0.016 & \textbf{0.867 $\pm$ 0.020} & 0.911 $\pm$ 0.020 & 0.906 $\pm$ 0.014 & \textbf{0.915 $\pm$ 0.021} \\
    CelebA   & 0.781 $\pm$ 0.017 & \textbf{0.901 $\pm$ 0.024} & 0.858 $\pm$ 0.019 & 0.706 $\pm$ 0.042 & \textbf{0.713 $\pm$ 0.042} & 0.698 $\pm$ 0.045 & 0.781 $\pm$ 0.032 & 0.828 $\pm$ 0.000 & \textbf{0.830 $\pm$ 0.028}  \\
    \bottomrule
    \end{tabular}
\end{table*}

\subsection{The Embedding Extractor}
\label{subsection:embedding_extractor}
Recall that all three target models contain an embedding extractor that maps an image to an embedding.
Tian et al.~\cite{TWKTI20} showed that in few-shot learning, the quality of the embedding extractor has some impact on the target model performance.
We now investigate the impact of the embedding extractor on the auditing performance for the following architectures: 

\begin{itemize}[noitemsep]
    \item \textbf{SimpleCNN} We adopt a two-layer convolutional neural network, which is the default feature extractor for other parts of the evaluation.
    \item \textbf{MobileNet}~\cite{SHZZC18} is an efficient model pretrained on the ImageNet dataset and widely adopted for mobile and embedded vision applications.
    \item \textbf{ResNet18}~\cite{HZRS16} is a public model with a deep residual network structure and pretrained on the ImageNet dataset.
    \item \textbf{ResNet50}~\cite{HZRS16} is a public model with a deeper residual network structure and pretrained on the ImageNet dataset. 
    \item \textbf{GoogleNet}~\cite{SLJSRAEVR15} is a convolutional neural network based on the Inception architecture, which allows the network to choose between multiple convolutional filter sizes in each block. 
\end{itemize}

\autoref{table:feature_extractor} illustrates the experimental results.
In general, we observe that the embedding extractor can help improve the target model performance but only slightly impacts the auditing performance in most settings. 

\begin{table*}[!ht]
    \centering
    \caption{Evaluation on different feature extractors \feat on four datasets and three target models. 
    }
    \vspace{-0.2cm}
    \label{table:feature_extractor}
    \setlength{\tabcolsep}{0.5em}
    \renewcommand{\arraystretch}{1.1}
    \footnotesize
    \begin{tabular}{c| c | c | c | c | c | c | c }
    \toprule
    & \multicolumn{1}{c|}{\textbf{Target Model}} 
    & \multicolumn{2}{c|}{\textbf{\siamese}}
    & \multicolumn{2}{c|}{\textbf{\proto}} 
    & \multicolumn{2}{c}{\textbf{\relation}} \\
    \multicolumn{1}{c|}{\textbf{Dataset}} & \multicolumn{1}{c|}{\textbf{Feature Extractor \feat }} & \textbf{$\model_{Target}$ Acc.} & \textbf{$\model_{Auditor}$ AUC} & \textbf{$\model_{Target}$ Acc.} & \textbf{$\model_{Auditor}$ AUC} & \textbf{$\model_{Target}$ Acc.} & \textbf{$\model_{Auditor}$ AUC} \\
    \toprule 
       & SimpleCNN & 0.568 & 0.995 $\pm$ 0.002 & 0.806 & 0.853 $\pm$ 0.041 & 0.856 & \textbf{0.900 $\pm$ 0.021} \\
       & MobileNet & 0.593 & \textbf{0.999 $\pm$ 0.001} & 0.699 & 0.823 $\pm$ 0.047 & 0.748 & 0.779 $\pm$ 0.032 \\
       & GoogleNet & 0.535 & 0.993 $\pm$ 0.003 & 0.710 & \textbf{0.885 $\pm$ 0.033} & 0.735 & 0.839 $\pm$ 0.059 \\
       & ResNet18 & 0.593 & 0.992 $\pm$ 0.003 & 0.729 & 0.850 $\pm$ 0.028 & 0.848 & 0.862 $\pm$ 0.028 \\
      \multirow{-5}{*}{\rotatebox[origin=c]{90}{UMDFaces}}
       & ResNet50 & 0.595 & 0.996 $\pm$ 0.002 & 0.733 & 0.830 $\pm$ 0.021 & 0.782 & 0.884 $\pm$ 0.030 \\
       \midrule 
       & SimpleCNN & 0.398 & 0.956 $\pm$ 0.015 & 0.583 & 0.876 $\pm$ 0.023 & 0.711 & 0.900 $\pm$ 0.022 \\
       & MobileNet & 0.545 & \textbf{0.994 $\pm$ 0.007} & 0.541 & \textbf{0.890 $\pm$ 0.021} & 0.786 & \textbf{0.907 $\pm$ 0.033} \\
       & GoogleNet & 0.485 & 0.983 $\pm$ 0.008 & 0.568 & 0.843 $\pm$ 0.037 & 0.725 & 0.873 $\pm$ 0.051 \\
       & ResNet18 & 0.515 & 0.989 $\pm$ 0.006 & 0.593 & 0.847 $\pm$ 0.027 & 0.688 & 0.869 $\pm$ 0.043 \\
      \multirow{-5}{*}{\rotatebox[origin=c]{90}{WebFace}}
       & ResNet50 & 0.477 & 0.991 $\pm$ 0.007 & 0.548 & 0.793 $\pm$ 0.034 & 0.694 & 0.798 $\pm$ 0.036 \\
       \midrule 
       & SimpleCNN & 0.662 & 0.998 $\pm$ 0.002 & 0.882 & 0.867 $\pm$ 0.020 & 0.934 & 0.915 $\pm$ 0.021 \\
       & MobileNet & 0.685 & 0.993 $\pm$ 0.001 & 0.868 & 0.867 $\pm$ 0.025 & 0.917 & 0.915 $\pm$ 0.011 \\
       & GoogleNet & 0.640 & \textbf{0.998 $\pm$ 0.004} & 0.877 & \textbf{0.870 $\pm$ 0.026} & 0.903 & 0.921 $\pm$ 0.039 \\
       & ResNet18 & 0.677 & 0.995 $\pm$ 0.002 & 0.903 & 0.845 $\pm$ 0.034 & 0.935 & 0.892 $\pm$ 0.024 \\
      \multirow{-5}{*}{\rotatebox[origin=c]{90}{VGGFace2}}
       & ResNet50 & 0.637 & 0.997 $\pm$ 0.004 & 0.923 & 0.852 $\pm$ 0.023 & 0.952 & \textbf{0.933 $\pm$ 0.036} \\
       \midrule 
       & SimpleCNN & 0.522 & \textbf{0.858 $\pm$ 0.019} & 0.837 & 0.698 $\pm$ 0.045 & 0.886 & 0.830 $\pm$ 0.028 \\
       & MobileNet & 0.590 & 0.812 $\pm$ 0.022 & 0.793 & \textbf{0.721 $\pm$ 0.052} & 0.872 & 0.853 $\pm$ 0.048 \\
       & GoogleNet & 0.562 & 0.777 $\pm$ 0.020 & 0.764 & 0.700 $\pm$ 0.038 & 0.845 & \textbf{0.871 $\pm$ 0.081} \\
       & ResNet18 & 0.585 & 0.796 $\pm$ 0.020 & 0.804 & 0.690 $\pm$ 0.052 & 0.867 & 0.834 $\pm$ 0.033 \\
      \multirow{-5}{*}{\rotatebox[origin=c]{90}{CelebA}}
       & ResNet50 & 0.555 & 0.785 $\pm$ 0.014 & 0.869 & 0.700 $\pm$ 0.050 & 0.902 & 0.845 $\pm$ 0.018 \\
    \bottomrule
    \end{tabular}
\end{table*}

\begin{table*}[!ht]
    \centering
    \caption{
    Auditing performance under input perturbation on UMDFaces. 
    The higher the perturbation level, the better the privacy-preserving level. 
    To give a direct impression of the perturbation, we show a visualization of different perturbation levels in \autoref{figure:input_perturbation_images}.
    }
    \vspace{-0.2cm}
    \label{table:input_perturbation}
    \setlength{\tabcolsep}{0.8em}
    \renewcommand{\arraystretch}{1.1}
    \footnotesize
    \begin{tabular}{c | c | c | c | c | c | c }
    \toprule
    \multicolumn{1}{c|}{\textbf{Target Model}} 
    & \multicolumn{2}{c|}{\textbf{\siamese}}
    & \multicolumn{2}{c|}{\textbf{\proto}} 
    & \multicolumn{2}{c}{\textbf{\textbf{\relation}}} \\
    \textbf{Perturbation Level} & $\model_{Target}$ Acc. ($\Delta$) & $\model_{Auditor}$ Acc. & $\model_{Target}$ Acc. ($\Delta$) & $\model_{Auditor}$ Acc. & $\model_{Target}$ Acc. ($\Delta$) & $\model_{Auditor}$ Acc. \\
    \toprule
    \rowcolor{mygray}
    \textbf{Original} & \textbf{0.500} & \textbf{0.991 $\pm$ 0.000} & \textbf{0.782} & \textbf{0.879 $\pm$ 0.000} & \textbf{0.847} & \textbf{0.961 $\pm$ 0.000} \\
    Low & 0.485 (-0.015) & 1.000 $\pm$ 0.001 & 0.803 (+0.021) & 0.785 $\pm$ 0.073 & 0.874 (+0.027) & 0.914 $\pm$ 0.019 \\
    Middle & 0.496 (-0.004) & 0.993 $\pm$ 0.002 & 0.843 (+0.061) & 0.852 $\pm$ 0.032 & 0.877 (+0.030) & 0.903 $\pm$ 0.017 \\
    High & 0.477 (-0.023) & 0.973 $\pm$ 0.004 & 0.777 (-0.005) & 0.843 $\pm$ 0.027 & 0.838 (-0.009) & 0.913 $\pm$ 0.021 \\
    \bottomrule
    \end{tabular}
\end{table*}

\begin{figure*}[!ht]
    \centering
    \begin{subfigure}{1.6\columnwidth}
    \includegraphics[width=\textwidth]{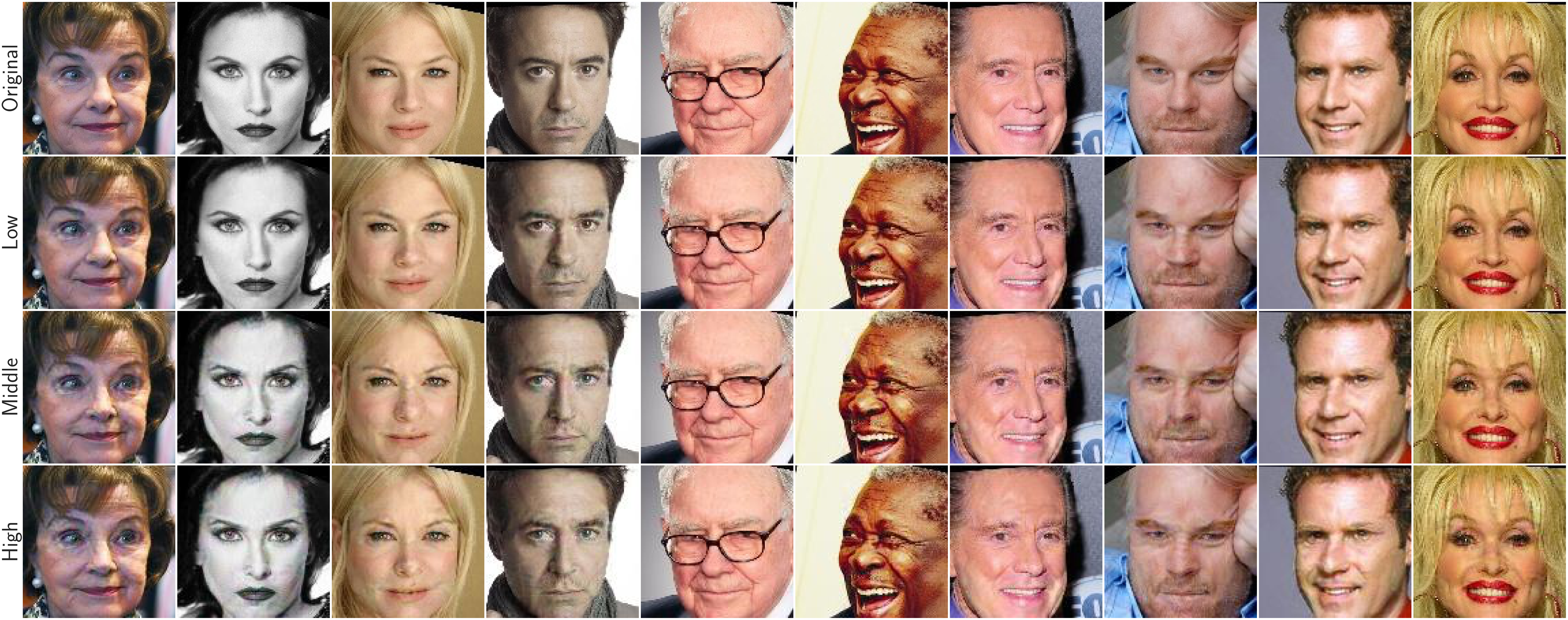}
    \label{subfigure:fawkes_perturbed}
    \end{subfigure}
    \vspace{-0.5cm}
    \caption{
    An illustration of images under different levels of adversarial noise perturbation.}
    \label{figure:input_perturbation_images}
\end{figure*}

\section{Robustness of \sysname}
\label{app:robustness}
In this section, we investigate the robustness of \sysname when the target models' pipeline is perturbed to evade auditing.
Concretely, we consider four defense mechanisms: \textit{Input perturbation} in \autoref{subsection:input_perturbation} (perturb the training images), \textit{training perturbation} in \autoref{subsection:train_perturbation} (perturb the training gradient by enforcing differential privacy), and \textit{output perturbation} in \autoref{subsection:output_perturbation} (perturb the similarity scores returned by the target models). 
In the end, we also explore an adaptive adversary scenario in \autoref{subsection:memguard}.

\subsection{Input Perturbation}
\label{subsection:input_perturbation}

\mypara{Methodology}
Multiple techniques are proposed to perturb the face images before training the facial recognition models~\cite{CGTFJB21,ESK21,WZWHBCZ23} and prevent the face images from being misused.
In our experiments, we consider a recently proposed technique called Fawkes~\cite{SWZLZZ20}.
The general idea of Fawkes is to add imperceptible noise to the target images that drive the embeddings of the face images to deviate from that of the raw face images.
According to its homepage, it has been downloaded more than $840,000$ times and used in various applications. 

\mypara{Visualization of Adversarial Perturbation}
\label{app:supplementary_fawkes_images}
We visualize training images under three input adversarial perturbation levels by the Fawkes~\cite{SWZLZZ20} in \autoref{figure:input_perturbation_images}.

\mypara{Evaluation}
The open-sourced Fawkes implementation\footnote{\url{https://github.com/Shawn-Shan/fawkes}} allows us to choose three perturbation levels: Low, middle, and high.
We experiment on the UMDFaces dataset and three target model architectures.
Concretely, we first use Fawkes with three perturbation levels to prepossess all the images in UMDFaces, and then use the same pipeline introduced in \autoref{subsection:train_phase} to build our shadow model and auditing model.

We report the target model performance and the auditing performance under different perturbation levels in \autoref{table:input_perturbation}.
We observe a slight performance drop of the target model when applying high-level perturbation, indicating the perturbed face images (especially under high-level perturbation) are more difficult to train. 
Regarding the auditing performance, we only observe a slight drop (the drop percentage is less than $6\%$), which indicates that \textbf{\sysname is robust to input perturbation}.

\begin{table*}[t]
    \centering
    \caption{
    Auditing performance under training perturbation. 
    We report the target model performance and auditing performance for three different privacy-preserving levels, i.e., Low, Middle, and High.
    Original means the target model without enforcing DP-SGD.
    The privacy budgets for the three privacy levels are $1.02$, $3.32$, and $47.35$.
    }
    \vspace{-0.2cm}
    \label{table:training_perturbation}
    \setlength{\tabcolsep}{0.5em}
    \renewcommand{\arraystretch}{1.1}
    \footnotesize
    \begin{tabular}{c| c | c | c | c | c | c | c }
    \toprule
    & \multicolumn{1}{c|}{\textbf{Target Model}} 
    & \multicolumn{2}{c|}{\textbf{\siamese}}
    & \multicolumn{2}{c|}{\textbf{\proto}} 
    & \multicolumn{2}{c}{\textbf{\relation}} \\
    \multicolumn{1}{c|}{\textbf{Dataset}} & \multicolumn{1}{c|}{\textbf{Perturbation Level}} & \textbf{$\model_{Target}$ Acc.} & \textbf{$\model_{Auditor}$ AUC} & \textbf{$\model_{Target}$ Acc.} & \textbf{$\model_{Auditor}$ AUC} & \textbf{$\model_{Target}$ Acc.} & \textbf{$\model_{Auditor}$ AUC} \\
    \toprule
       \rowcolor{mygray}
       \cellcolor{white}
       & \textbf{Original} & \textbf{0.500} & \textbf{0.996 $\pm$ 0.003} & \textbf{0.793} & \textbf{0.883 $\pm$ 0.022} & \textbf{0.852} & \textbf{0.935 $\pm$ 0.012} \\
       & Low & 0.357 & 0.918 $\pm$ 0.009 & 0.454 & 0.919 $\pm$ 0.000 & 0.212 & 0.941 $\pm$ 0.018 \\
       & Middle & 0.370 & 0.941 $\pm$ 0.010 & 0.455 & 0.905 $\pm$ 0.024 & 0.211 & 0.938 $\pm$ 0.021 \\
      \multirow{-4}{*}{\rotatebox[origin=c]{90}{UMDFaces}}
       & High & 0.273 & 0.942 $\pm$ 0.008 & 0.490 & 0.902 $\pm$ 0.017 & 0.201 & 0.947 $\pm$ 0.012 \\
       \midrule 
       \rowcolor{mygray}
       \cellcolor{white}
       & \textbf{Original} & \textbf{0.390} & \textbf{0.986 $\pm$ 0.009} & \textbf{0.607} & \textbf{0.875 $\pm$ 0.028} & \textbf{0.756} & \textbf{0.905 $\pm$ 0.023} \\
       & Low & 0.287 & 0.960 $\pm$ 0.008 & 0.364 & 0.940 $\pm$ 0.000 & 0.200 & 0.944 $\pm$ 0.000 \\
       & Middle & 0.242 & 0.960 $\pm$ 0.018 & 0.370 & 0.920 $\pm$ 0.005 & 0.202 & 0.939 $\pm$ 0.000 \\
      \multirow{-4}{*}{\rotatebox[origin=c]{90}{WebFace}}
       & High & 0.235 & 0.938 $\pm$ 0.010 & 0.353 & 0.911 $\pm$ 0.000 & 0.196 & 0.924 $\pm$ 0.015 \\
       \midrule 
       \rowcolor{mygray}
       \cellcolor{white}
       & \textbf{Original} & \textbf{0.575} & \textbf{0.996 $\pm$ 0.002} & \textbf{0.868} & \textbf{0.866 $\pm$ 0.016} & \textbf{0.943} & \textbf{0.906 $\pm$ 0.014} \\
       & Low & 0.330 & 0.981 $\pm$ 0.004 & 0.433 & 0.929 $\pm$ 0.000 & 0.215 & 0.932 $\pm$ 0.011 \\
       & Middle & 0.250 & 0.990 $\pm$ 0.007 & 0.425 & 0.877 $\pm$ 0.024 & 0.214 & 0.913 $\pm$ 0.019 \\
      \multirow{-4}{*}{\rotatebox[origin=c]{90}{VGGFace2}}
       & High & 0.258 & 0.982 $\pm$ 0.005 & 0.405 & 0.885 $\pm$ 0.024 & 0.215 & 0.909 $\pm$ 0.012 \\
       \midrule 
       \rowcolor{mygray}
       \cellcolor{white}
       & \textbf{Original} & \textbf{0.435} & \textbf{0.901 $\pm$ 0.024} & \textbf{0.812} & \textbf{0.713 $\pm$ 0.042} & \textbf{0.867} & \textbf{0.828 $\pm$ 0.000} \\
       & Low & 0.333 & 0.804 $\pm$ 0.020 & 0.355 & 0.727 $\pm$ 0.037 & 0.211 & 0.849 $\pm$ 0.000 \\
       & Middle & 0.325 & 0.795 $\pm$ 0.028 & 0.430 & 0.751 $\pm$ 0.020 & 0.205 & 0.870 $\pm$ 0.000 \\
      \multirow{-4}{*}{\rotatebox[origin=c]{90}{CelebA}}
       & High & 0.242 & 0.789 $\pm$ 0.011 & 0.390 & 0.623 $\pm$ 0.000 & 0.208 & 0.891 $\pm$ 0.000 \\
    \bottomrule
    \end{tabular}
\end{table*}

\subsection{Training Perturbation}
\label{subsection:train_perturbation}
\mypara{Methodology}
A generic approach to protect users' data privacy is differential privacy (DP), which guarantees that any sample in the input dataset has a limited impact on the final output~\cite{ZWHLBHCZ21,DZBLJCC21,ZWLHC18,WCZSCLLJ21,DHZFCZG23,WZWHBCZ23,YZDCCS23}.
For machine learning models, the most representative DP algorithm is Differentially-Private Stochastic Gradient Descent (DP-SGD)~\cite{ACGMMTZ16}.
In general, DP-SGD adds Gaussian noise to gradient $g$ during the target ML model's training process, i.e., $\Tilde{g} = g + \mathcal{N}\left(0, \Delta_g^2 \sigma^2 \mathbf{I} \right)$.
Note that there is no prior knowledge to determine the influence of a single training sample on the gradient $g$; thus, the sensitivity of $g$ cannot be directly computed.
To address this problem, DP-SGD proposes to bound the $\ell_2$ norm of the gradient to $C$ by clipping $g$ to $g/\max\{1, ||g||_2 / C\}$.
This clipping ensures that if $||g||_2 \leq C$, $g$ is preserved; otherwise, it gets scaled down to be the norm of $C$.
As such, the sensitivity of $g$ is bounded by $C$.
Note that we aim to show the defensive performance of adding perturbation in the training process of the target model. 
Besides, existing user-level DP mainly focus on the federated learning setting~\cite{MRTZ18,MSCS19}. 
They do not fit to few-shot learning paradigms. 

\mypara{Evaluation}
We conduct experiments on four datasets and three target models.
The experimental results in \autoref{table:training_perturbation} show that \textbf{DP-SGD has a severe impact on the target model performance}.

We further observe variations in the auditing performance across the three model architectures: 
\siamese is more sensitive to DP-SGD while \proto and \relation are less sensitive.
Take VGGFace2 as an example. Applying a high-level noise to the training phase of \siamese makes target model accuracy drop by $55\%$, while the auditing accuracy only drops $1.4\%$.
On the other hand, the auditing accuracy of \proto and \relation remains almost the same.

\subsection{Output Perturbation}
\label{subsection:output_perturbation}
\mypara{Methodology}
Another approach to protecting ML models from inference attacks is adding perturbations to the target models' outputs.
In this subsection, we evaluate the robustness of our auditing model when the similarity scores returned by the target model are perturbed.
We implement this defense by adding a zero-mean Laplace noise with a standard deviation $\delta$ to the target model's outputs.

\begin{figure*}[!tpb]
    \centering
    \begin{subfigure}{2\columnwidth}
    \includegraphics[width=\textwidth]{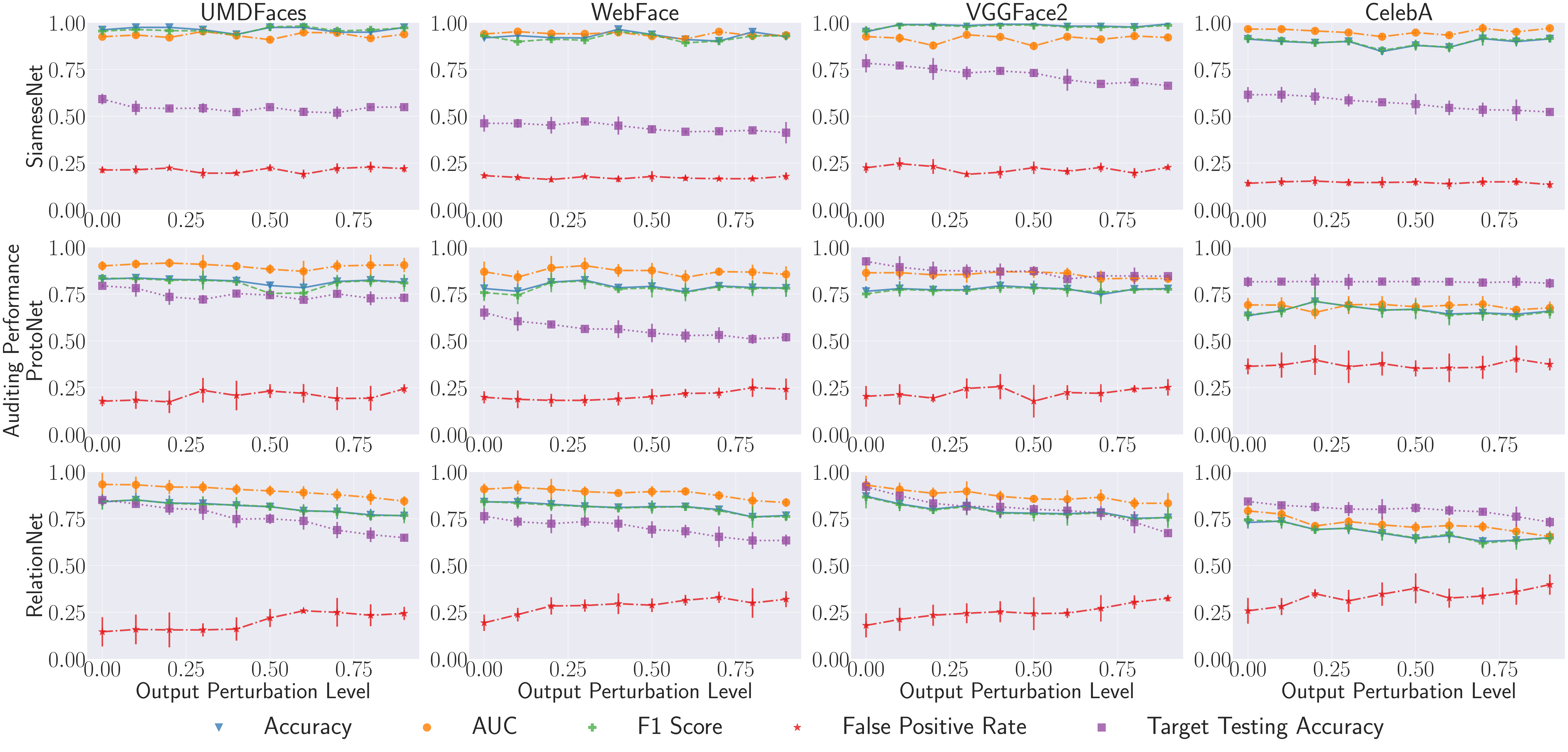}    
    \label{subfigure:output_perturbation_defense}
    \end{subfigure}
    \vspace{-0.8cm}
    \caption{Auditing performance under output perturbations. 
    The x-axis represents different noise levels used to perturb the target model's outputs. 
    Higher values mean a stronger perturbation degree. The y-axis represents the auditing performance.
    }
    \label{figure:output_perturbation_defense}
\end{figure*}

\mypara{Evaluation}
We conduct experiments on all four datasets and three target models.
The experimental results in \autoref{figure:output_perturbation_defense} show that \textbf{\sysname is robust to output perturbation}.
Concretely, the auditing performance on \siamese and \proto does not drop significantly, and the auditing performance drop on \relation is in the scope of $15\%$.
We also observe a slight drop in the target model performance when the noise perturbation level increases, which indicates the robustness of \sysname.

\begin{figure*}[!ht]
    \centering
    \begin{subfigure}{1.8\columnwidth}
    \includegraphics[width=\textwidth]{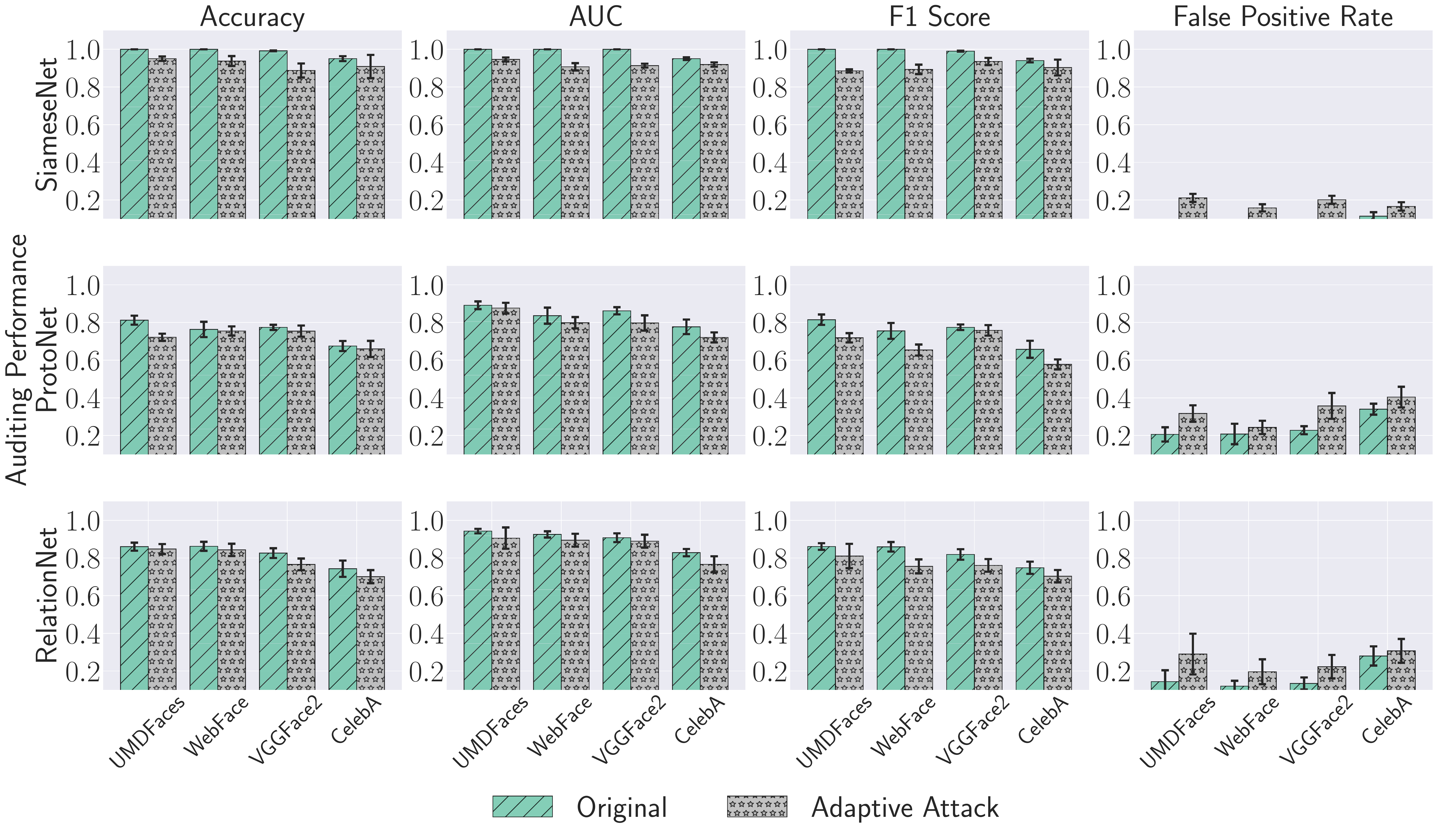}
    \label{subfigure:memguard}
    \end{subfigure}
    \vspace{-0.8cm}
    \caption{Auditing performance comparison under an adaptive adversary against \sysname.}
    \label{figure:memguard}
\end{figure*}

\subsection{MemGuard}
\label{subsection:memguard}
\mypara{Threat Model}
In practice, a malicious data collector might be aware of the existence of \sysname. They modify their facial recognition models in a way to evade auditing and gain financial benefits or avoid a lawsuit.
We evaluate the performance of \sysname when the target model's output is perturbed to avoid membership auditing.

\mypara{Methodology}
We follow the design intuition of MemGuard~\cite{JSBZG19} and perform adaptive attacks against \sysname.
The general idea is to perturb the similarity scores (outputs of the target model) while achieving two objectives: Minimum label loss and maximum auditing confusion.
The first goal guarantees the noisy posteriors do not change the predicted labels of the target model given any inputs.
The second goal aims to make \sysname randomize its predictions of the user-level membership status given any face images to be audited.
Concretely, to make \sysname unable to distinguish member and non-member users, the adaptive attacker aims to add the maximum noise on the similarity scores under the constraint of not affecting the corresponding label.
To ensure that the final summation of the target model's output is valid (summing to one), after adding the maximum noise to the target similarity score, we apply a SoftMax function to the entire similarity score vector, generating the final perturbed score vector.
Note that the adversary cannot perturb the reference information as it is prepared by \sysname and is a fixed value given any input images; thus, we concatenate the original reference information and the perturbed similarity scores as the auditing feature.

\mypara{Results}
\autoref{figure:memguard} illustrates the auditing performance of \sysname under an adaptive attack.
We observe that adaptive perturbation on the target model's outputs only slightly affects the auditing performance.
The drop in auditing performance is less than $5\%$.
This differs from the sample-level membership inference case, in which MemGuard leads to near-random guessing attack performance.
There are three reasons.
First, MemGuard can only perturb one value of the auditing feature per query, while \sysname queries the target model multiple times and combine the similarity scores of multiple queries as the auditing feature. 
Second, in sample-level membership inference, an adaptive adversary can perturb the whole attack feature (the posterior of the target sample) simultaneously, but it can only perturb one value per query in our user-level membership inference setting.
Third, the reference information helps maintain the relative correlation of the query images and captures the subtle difference between member and non-member users.
Additionally, our experimental results (in \autoref{subsection:output_perturbation}) show a limited impact of output perturbation even without the minimum label loss constraint. 
In summary, the similarity scores are more difficult for an adaptive adversary to perturb than output perturbation due to the minimum label loss constraint, which keeps the predicted label of a query sample fixed for a given support set and leaves the adversary little room to perturb. 

\end{document}